\documentclass[aps,pra,nobalancelastpage,superscriptaddress, twocolumn, 10pt]{revtex4-2}

\usepackage{color,ifthen,amsthm,amsmath,amsxtra,amsfonts,dsfont,graphicx,bm,tikz,scalerel,wasysym,bbm,graphicx,amsthm, braket,physics,empheq}
\usepackage[colorlinks=true,linkcolor=blue, citecolor=blue, urlcolor=blue, bookmarks]{hyperref}
\usepackage[inline]{enumitem}

\usepackage{tikz}
\usetikzlibrary{hobby,arrows,arrows.meta,calc,matrix,backgrounds,external,fit,positioning,decorations.pathmorphing,decorations.markings,patterns,decorations.pathreplacing,decorations.text,shapes.misc,external,mindmap}

\definecolor{colUc}{rgb}{0.71,0.41,0.42}
\definecolor{colU}{rgb}{0.71,0.8,0.76}
\definecolor{colLines}{rgb}{0.31,0.31,0.31}
\definecolor{colIstC}{rgb}{0.73,0.69,0.7}
\definecolor{colSt}{rgb}{0.96,0.74,0.59}
\definecolor{colIst}{rgb}{0.81,0.77,0.78}

\newcommand\prop[3]{
    \draw[thick,colLines,fill=#3,rounded corners=1]
    ({((#1)-0.3)},{((#2)-0.3)}) rectangle ({((#1)+0.3)},{((#2)+0.3)});
}

\newcommand\istate[3]{
    \draw [thick,rounded corners=1,colLines,fill=#3] ({(#1)},{(#2)}) +(0.1,0.1) rectangle 
    +(-0.1,-0.1);
}

\newcommand{\be}{\begin{equation}}
\newcommand{\ee}{\end{equation}}
\newcommand{\bea}{\begin{align}}
\newcommand{\eea}{\end{align}}
\newcommand{\bse}{\begin{subequations}}
\newcommand{\ese}{\end{subequations}}

\theoremstyle{plain}
\newcommand{\1}{\mathbbm{1}}

\theoremstyle{plain}

\theoremstyle{plain}

\begin{document}

\title{Growth of R\'enyi entropies in interacting integrable models and the breakdown of the quasiparticle picture}

\author{Bruno Bertini}
\affiliation{School of Physics and Astronomy, University of Nottingham, Nottingham, NG7 2RD, UK}
\affiliation{Centre for the Mathematics and Theoretical Physics of Quantum Non-Equilibrium Systems,
University of Nottingham, Nottingham, NG7 2RD, UK}

\author{Katja Klobas}
\affiliation{Rudolf Peierls Centre for Theoretical Physics, Clarendon Laboratory, Oxford OX1 3PU, UK}

\author{Vincenzo Alba}
\affiliation{Dipartimento di Fisica, Universit\`a di Pisa, and INFN Sezione di Pisa, Largo Bruno Pontecorvo 3, Pisa, Italy}

\author{Gianluca Lagnese}
\affiliation{SISSA and INFN Sezione di Trieste, via Bonomea 265, 34136 Trieste, Italy}

\author{Pasquale Calabrese}
\affiliation{SISSA and INFN Sezione di Trieste, via Bonomea 265, 34136 Trieste, Italy}
\affiliation{International Centre for Theoretical Physics (ICTP), Strada Costiera 11, 34151 Trieste, Italy}

\begin{abstract}

R\'enyi entropies are conceptually valuable and experimentally relevant generalisations of the celebrated von Neumann entanglement entropy. After a quantum quench in a clean quantum many-body system they generically display a universal linear growth in time followed by saturation. While a finite subsystem is essentially at local equilibrium when the entanglement saturates, it is genuinely out-of-equilibrium in the growth phase. In particular, the slope of the growth carries vital information on the nature of the system's dynamics, and its characterisation is a key objective of current research. Here we show that the slope of R\'enyi entropies can be determined by means of a spacetime duality transformation. In essence, we argue that the slope coincides with the stationary density of entropy of the model obtained by exchanging the roles of space and time. Therefore, very surprisingly, the slope of the entanglement is expressed as an equilibrium quantity. We use this observation to find an explicit exact formula for the slope of R\'enyi entropies in all integrable models treatable by thermodynamic Bethe ansatz and evolving from integrable initial states. Interestingly, this formula can be understood in terms of a quasiparticle picture only in the von Neumann limit.
\end{abstract}

\maketitle

\section{Introduction}
The linear growth of entanglement is arguably the most distinctive and pervasive phenomenon observed in the context of quantum many-body dynamics: whenever a clean, locally interacting many-body system is prepared in a low-entangled state and then let to evolve, the entanglement between a compact region and its complement grows linearly in time. The ubiquity of this phenomenon suggests that a universal underlying mechanism is hidden behind the scenes. An astonishing outcome of recent research, however, suggests that this is not the case. Two distinct mechanisms for entanglement growth have been identified depending on the nature of the dynamics. 

The first account of linear growth of entanglement has been given in the context of (1+1)-dimensional conformal field theory~\cite{calabrese2005evolution}, where it was explained assuming that the entanglement is ``spread'' throughout the system by pairs of correlated quasiparticles produced by the quench~\cite{calabrese2005evolution}. This intuitive quasiparticle picture has then been extended to quantitatively characterise the dynamics of the standard measure of bipartite entanglement --- the von Neumann entanglement entropy or simply entanglement entropy~\cite{amico2008entanglement,calabrese2009introduction,laflorencie2016quantum} -- in many different kinds of systems with stable quasiparticles, such as free~\cite{calabrese2005evolution, fagotti2008evolution, alba2018entanglement} and interacting integrable models~\cite{alba2017entanglement, alba2018entanglement,lagnese2022entanglement} in a large variety of physical contexts. Few years later, however, the same phenomenology has been observed in systems with no quasiparticles, for instance holographic conformal field theories~\cite{liu2014entanglement, asplund2015entanglement}, generic interacting systems \cite{laeuchli2008spreading, kim2013ballistic}, and chaotic quantum circuits~\cite{pal2018entangling, bertini2019entanglement, piroli2020exact, gopalakrishnan2019unitary}. This unexpected ballistic growth of entanglement in chaotic systems has finally been explained through a ``minimal membrane'' picture~\cite{nahum2017quantum, zhou2020entanglement}. In essence, the idea is that in chaotic systems the entanglement between two complementary regions is measured by the tension of the minimal spacetime surface that separates the two. Even though both quasiparticle and minimal-membrane pictures explain the linear growth of entanglement, they predict qualitatively different phenomenology when considering more complicated partitions of the system~\cite{asplund2015entanglement, nahum2017quantum, alba2019quantum} or for finite sizes~\cite{nahum2017quantum, bertini2019entanglement}.

Such a twofold description of entanglement growth, however, has recently been challenged by studies on the dynamics of R\'enyi entropies. These are a family of seemingly minor variations of the entanglement entropy (see the definition below), which have been shown to provide highly nontrivial universal information about the system~\cite{laflorencie2016quantum}, for instance on its topological properties~\cite{li2008entanglement}. Arguably the most significant point of interest of R\'enyi entropies, of integer order $\alpha\geq2$, is that they are accessible in present-day experiments~\cite{Greiner2015,Greiner2016,Linke2018,Greiner2019,elben2020mixed,zhou2020single,neven2021symmetry}. 

Given the unquestionable success of the quasiparticle picture in quantitatively capturing the evolution of the entanglement entropy for integrable systems, it is very natural to assume that the same picture also describes the evolution of R\'enyi entropies. Several basic facts support this scenario: \begin{enumerate*}[label=(\roman*)] \item it holds for free systems~\cite{calabrese2005evolution, fagotti2008dinamica,fagotti2008evolution}, \item there is no clear qualitative difference between the numerically computed R\'enyi entropies and the entanglement entropy~\cite{alba2017quench, alba2017renyi}, \item in chaotic systems the membrane picture describes both von Neumann and R\'enyi entropies~\cite{nahum2017quantum, zhou2020entanglement}.\end{enumerate*} The extension of the quasiparticle picture to describe R\'enyi entropies in the presence of interactions, however, proved to be very challenging~\cite{alba2017quench, alba2017renyi,mestyan2018renyi, klobas2021exact}. In fact, Ref.~\cite{klobas2021entanglement} showed that no consistent quasiparticle picture can describe the evolution of R\'enyi entropies in an integrable quantum cellular automaton. A possible explanation of these findings is that, although the quasiparticle picture describes the evolution of entanglement entropy also in the presence of interactions, it fails to describe the growth of R\'enyi entropies. This would highlight a very unexpected fundamental difference between the two quantities, which complements the accounts of sub-linear growth of R\'enyi entropies in certain systems with diffusive conservation laws~\cite{rakovszky2019sub, huang2020dynamics,znidaric2020entanglement}.

Motivated by this question here we investigate the dynamics of R\'enyi entropies in one-dimensional quantum many-body systems using a radically different approach. Our main idea is to argue that, if the roles of space and time are exchanged, the slope of a given R\'enyi entropy is mapped to the density of the same entropy in an appropriate steady state. This essentially means that the exchange of space and time --- which we dub ``spacetime swap'' --- maps the calculation of a non-equilibrium quantity into that of an equilibrium one. 

To demonstrate the validity of aforementioned correspondence under spacetime swap we begin considering locally interacting systems in discrete space time: the so called local quantum circuits. Indeed, as recently pointed out in Ref.~\cite{ippoliti2021fractal}, in these systems the correspondence can be established rigorously using the spacetime duality method introduced in Ref.~\cite{bertini2019entanglement} (see also \cite{piroli2020exact,klobas2021entanglement} for further developments). In particular, for  dual-unitary circuits~\cite{bertini2019exact}, where the dynamics from a class of compatible initial states~\cite{bertini2019entanglement, piroli2020exact} are essentially invariant under the exchange of space and time, one has that the slope of a given R\'enyi entropy slope coincides with the entropy density of the infinite-temperature state --- the stationary state of the space evolution.  

Then we consider another class of systems where the dynamics are essentially invariant under an appropriate spacetime swap: relativistic quantum field theories. In this case we show that the correspondence holds in the free case. Assuming that it continues to hold for interacting integrable quantum field theories~\cite{zamolodchikov1979factorised, zamolodchikov1990thermodynamic, mussardo2010statistical} when evolving from appropriate compatible initial states~\cite{ghoshal1994boundary}, we arrive at a formula for the slope of R\'enyi entropies in all such systems. 

Finally we extend our result to {\it all} interacting integrable models treatable by thermodynamic Bethe ansatz (TBA)~\cite{yang1969thermodynamics, korepin1997quantum, takahashi1999thermodynamics, zamolodchikov1990thermodynamic, tongeren2016introduction} and evolving from compatible initial states~\cite{piroli2017what}. A significant physical insight of our result is that, as for the special case of Ref.~\cite{klobas2021entanglement}, the dynamics of R\'enyi entropies cannot be understood in terms of a consistent quasiparticle picture.

The rest of this manuscript is laid out as follows. In Sec.~\ref{sec:setting} we define more precisely the setting considered and the quantities relevant for our analysis. In Sec.~\ref{sec:duality} we demonstrate the correspondence of slopes and densities of R\'enyi entropies under spacetime swap in local quantum circuits, where the discreteness of spacetime allows for a rigorous treatment. In Sec.~\ref{sec:qft} we discuss the case of relativistic quantum field theories. In particular, in Sec.~\ref{sec:iqft} we present our exact formula for the slope of all R\'enyi entropies in interacting integrable quantum field theories. In Sec.~\ref{sec:TBA} we derive a direct generalisation of this result to describe general TBA-integrable systems, and test it against exact analytical and numerical results. In Sec.~\ref{sec:quasiparticle} we consider the implications of our formula for the validity of the quasiparticle picture, and in Sec.~\ref{sec:finite} we discuss a possible extension to the finite-subsystem-regime. Finally, Sec.~\ref{sec:discussion} contains our conclusions.
A number of technical points are relegated to the appendices. 

\section{Setting} \label{sec:setting}

In this work we consider a quantum many-body system prepared in a non-equilibrium
initial state, $\ket{\Psi_0}$, which is pure and has low entanglement. At time
$t=0$ we let the system evolve under its own unitary dynamics, so that the
state at time $t>0$ is given by  
\be
\ket{\Psi_t} = \mathbb{U}^t \ket{\Psi_0},
\ee
where $\mathbb U$ is the time-evolution operator. As a result of the unitary
evolution the state becomes increasingly more entangled as time
advances~\cite{amico2008entanglement}. 
The entanglement between a finite region $A$ and the rest of the
system can be quantified computing the \emph{R\'enyi entropies}
\be
S^{(\alpha)}_A(t) = \frac{1}{1-\alpha}
\ln\left[\tr\rho^\alpha_A(t)\right], \qquad \alpha\in\mathbb R\,,
\label{eq:Srenyi}
\ee  
where $\rho_A(t)$ is the density matrix of the system reduced to the subsystem $A$.
In the limit $\alpha\to 1$ the above expression is reduced to the von Neumann entanglement
entropy
\be
S_A(t)=\lim_{\alpha\to 1} S_A^{(\alpha)}(t)=-\tr\left[\rho_{A}(t)\ln\rho_A(t)\right].
\ee
At times that are short compared to the subsystem size $\abs{A}$, R\'enyi entropies are expected to grow linearly (at least in the systems of interest here), while at sufficiently large times they saturate to their ``thermodynamic values'', i.e., they coincide with the R\'enyi entropies of the \emph{stationary state} describing the subsystem $A$. These two regimes can be respectively characterised by the \emph{entanglement slope} $s_{\alpha}$, and the \emph{stationary entanglement density}
$d_{\alpha}$. 

The density $d_{\alpha}$ is defined as the density of R\'enyi-$\alpha$ 
entanglement entropy of a finite subsystem of a thermodynamically large system in the
$t\to\infty$ limit, i.e.,
\be \label{eq:density}
    d_{\alpha}=\lim_{\abs{A}\to\infty}\lim_{t\to\infty}\left(\lim_{L\to\infty}
\frac{S_{A}^{(\alpha)}}{|{A}|}\right),
\ee
where $\abs{A}$ is the size of the subsystem $A$ and $L$ that of the total system. The final $|A|\to\infty$ limit is taken to remove the boundary effects and focus on the bulk physics of the subsystem $A$.

The asymptotic slope $s_{\alpha}$, sometimes also referred to as ``entanglement production rate'', is defined as the ratio of the R\'enyi-$\alpha$ entanglement between a large subsystem and the rest, and time
\be \label{eq:slope}
    s_{\alpha}=\lim_{t\to\infty}\lim_{\abs{A}\to\infty}\left(\lim_{L\to\infty}
\frac{S_{A}^{(\alpha)}}{2t}\right),
\ee
where the factor of $2$ accounts for the fact that the subsystem $A$ has two edges through which
it develops correlations with its complement. In analogy with the density, we
take the $t\to\infty$ limit to remove finite-time effects.

Our main goal is to establish a formal connection between the slope and the density based on a spacetime swap, i.e., an exchange of space and time.

\section{Spacetime swap for local quantum circuits} 
\label{sec:duality}
\begin{figure*}
    \includegraphics[width=1.5\columnwidth]{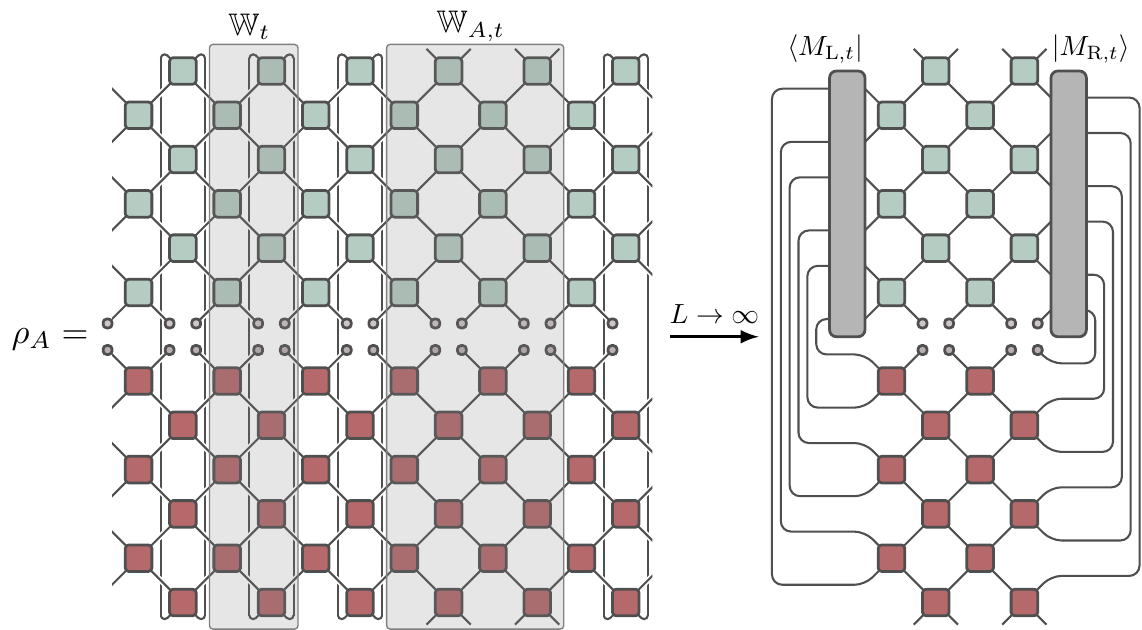}
    \caption{\label{fig:QC} Diagrammatic representation of the reduced density
    matrix.  The reduced density matrix $\rho_{A}$ is obtained by evolving an
    initial state in time and tracing over the complement $\overline{A}$ of the
    subsystem $A$ (graphically denoted by connecting the top and bottom legs).
    Alternatively, we can understand it as a trace of a product of powers of
    the space transfer matrix $\mathbb{W}_{t}$ and the transfer matrix
    $\mathbb{W}_{A,t}$ (both shaded in grey), as given by
    Eq.~\eqref{eq:rhoAsp}.  In the limit $L\to\infty$, the
    section of the tensor network corresponding to the rest of the system can
    be replaced by fixed-points $\bra{M_{{\rm L},t}}$ and $\ket{M_{{\rm R},t}}$
    of the space transfer matrix $\mathbb{W}_{t}$ (cf.\
    Eq.~\eqref{eq:rhoAfps}).}
\end{figure*}
A convenient setting for our analysis is that of \emph{local quantum circuits}. These are models in discrete space where the time evolution occurs in discrete steps through local updates. The discreteness of time-evolution and the locality of interactions imply that these systems are \emph{closed} under spacetime swap. Namely the ``dual system" obtained exchanging the roles of space and time in a quantum circuit is still a quantum circuit, although the unitarity of the time evolution is generically not preserved~\cite{ippoliti2021postselectionfree}. This simple observation gives a tool to analyse several properties of quantum circuits by ``evolution in space''~\cite{bertini2018exact,bertini2019exact,bertini2019entanglement,ippoliti2021postselectionfree,lu2021spacetime,garratt2021manybody}. In particular, as we now discuss, it can be used to write explicit expressions for the entanglement slope that closely resemble those for the density~\cite{bertini2019entanglement,klobas2021entanglement,bertini2022dynamics}. 

For the sake of clarity we consider \emph{brickwork} circuits --- i.e., circuits consisting of two-site gates applied to first even and then odd pairs of neighbouring sites --- but with minor modifications the argument can be repeated for any discrete-spacetime model with local unitary interactions.

More specifically we consider a chain of $2L$ sites, hosting qudits with $d$ internal states, and where the time evolution operator~$\mathbb{U}_{L}$ is written as the tensor product of $L$ two-site unitary gates $U$ multiplied by the same product shifted by one site, i.e.,
\be
\mathbb{U}_L=
\Pi_{2 L}^{\dagger} U^{\otimes L}\Pi_{2 L} U^{\otimes L},
\ee
where $\Pi_{2L}$ represents a periodic shift by one site on a chain of $2L$ sites. At time $t$ the reduced density matrix of a~subsystem $A$ is given by 
\be
\rho_{A}=\tr_{\bar{A}}\ketbra{\Psi_t}{\Psi_t}
=\tr_{\bar{A}}\left(\mathbb{U}^t \ketbra{\Psi_0}{\Psi_0}
\left.\mathbb{U}^{\dagger}\right.^{t}\right),
\ee
where we repeatedly apply the time-evolution operator $\mathbb{U}$ to the initial state, and then trace over the rest of the system $\bar{A}$. This can be represented graphically as a $2L\times 4t$ tensor network, see Fig.~\ref{fig:QC}. 

The same tensor network can be equivalently thought of as resulting from an evolution in \emph{space}. Indeed, rather than viewing gates as acting on the $2L$ sites arranged horizontally and propagating upwards/downwards, one can imagine them acting on the $4t$ sites arranged vertically and propagating rightwards/leftwards. To this end we introduce two different \emph{space transfer matrices} playing the role of evolution operators in space (cf.\ Fig.~\ref{fig:QC}): $\mathbb{W}_t$, that describes the space evolution in the complement of the subsystem $\overline{A}$, and $\mathbb{W}_{A,t}$, that acts both on $4t$ \emph{temporal} sites and on the  subsystem~$A$. 
Using these definitions, we can express the reduced density matrix $\rho_{A}$ as the trace of a large power of~$\mathbb{W}_t$ multiplied by $\mathbb{W}_{A,t}$, i.e.,
\be \label{eq:rhoAsp}
\rho_{A}=\tr\left(\mathbb{W}_t^{L-\abs*{{A}}} \mathbb{W}_{A,t}\right).
\ee
Due to the unitarity of the time-evolution (see, e.g.,~\cite{piroli2020exact,klobas2021exactrelaxation,banuls2009matrix,muellerhermes2012tensor,hastings2015connecting}) the transfer matrix $\mathbb{W}_t$ has a single non-degenerate eigenvalue $1$, while all the other eigenvalues are $0$. The latter generically correspond to nontrivial Jordan blocks of size smaller or equal to $t$. Therefore, when $L$ is larger than $t$, the matrix power 
\be
\mathbb{W}_t^{L-\abs*{{A}}}
\ee
can be replaced with a projector on the \emph{fixed points} $\ket{M_{{\rm L},t}}$ and $\ket{M_{{\rm R},t}}$, i.e., the left and right eigenvectors corresponding to eigenvalue 1~\cite{banuls2009matrix,muellerhermes2012tensor,hastings2015connecting}. Namely 
\be \label{eq:rhoAfps}
\lim_{L\to\infty}\rho_{A}=\mel{M_{{\rm L},t}}{\mathbb W_{A,t}}{M_{{\rm R},t}},
\ee
where we chose the normalisation such that 
\be
\braket{M_{{\rm L},t}}{M_{{\rm R},t}}=1\,.
\label{eq:fixedpointnormalization}
\ee
Eq.~\eqref{eq:rhoAfps} shows that the fixed points completely capture the
effect of $\bar A$ on the finite subsystem $A$ (cf.\ Fig.~\ref{fig:QC}). For
this reason, and to stress their connection with the Feynman-Vernon influence
functional, Ref.~\cite{lerose2021influence} (see
also~\cite{lerose2021scaling,sonner2021influence,sonner2022characterizing,giudice2022temporal})
proposed to dub them \emph{influence matrices}.

The fixed points $\ket{M_{{\rm L},t}}$ and $\ket{M_{{\rm R},t}}$ can be thought of either as vectors in the space of $4t$ temporal sites, or as matrices $M_{{\rm L},t}$, $M_{{\rm R},t}$ mapping from $2t$ temporal sites in the bottom half
to $2t$ sites in the top half~\footnote[2]{By this definition, $\bra{M_{{\rm L/R},t}}$
correspond to $M_{{\rm L/R},t}^{\ast}$, and $M_{{\rm L/R},t}^\dagger=M_{{\rm L/R},t}$.}
(see r.h.s.\ of Fig.~\ref{fig:QC} for an illustration).
\begin{figure*}
    \includegraphics[width=2\columnwidth]{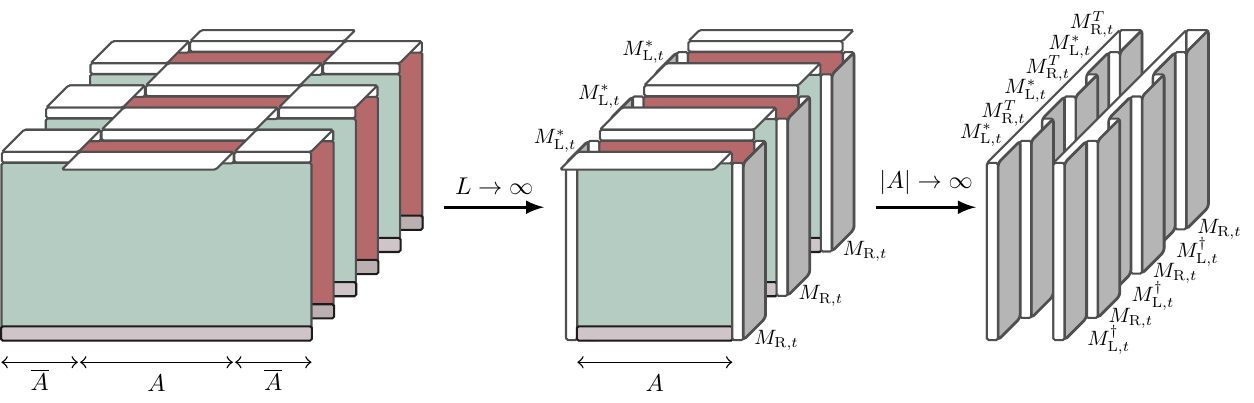}
    \caption{\label{fig:QCslope} Schematic illustration of $\tr\left[\rho_{A}^3\right]$.
    Green and red rectangles are condensed representation of the green and red
    half of the time-evolution from Fig.~\ref{fig:QC}. The white connections on
    the left represent trace over the rest of the system, while the connections
    on the right correspond to matrix products and then an overall trace.
    In the limit of large $L-\abs*{{A}}$ the left and right parts of $\overline{A}$
    can be substituted by fixed points $M_{\mathrm{L}, t}^{\ast}$ and
    $M_{\mathrm{R},t}$ connecting pairs of time-sheets. Analogously, when the subsystem
    size becomes large, the section in the middle can be replaced with fixed points
    $M_{\mathrm{L},t}^{\dagger}$, and $M_{\mathrm{R},t}^{T}$. Note the additional transpose
    of fixed points, which is the consequence of connecting the opposite parity of pairs
    of neighbours. Thus we obtain the right-most diagram, which corresponds precisely to
    the r.h.s.\ of Eq.~\eqref{eq:trPowfps}.
    }
\end{figure*}
The latter perspective makes fixed points convenient to access
$s_{\alpha}$.

To demonstrate this we begin by observing that, for $n$ integer, $\tr\rho_{A}^n$ can be represented as $n$ copies of conjugate pairs of the time-evolved initial state that are coupled in a staggered fashion. In the section corresponding to $\overline{A}$ the pairs are connected, while in $A$ the conjugate copy of a pair is connected to the non-conjugate copy of the next pair --- see the left panel of Fig.~\ref{fig:QCslope} for a pictorial representation. This means that $\tr\left[\rho_{A}^{n}(t)\right]$ can be expressed in terms of products of $n$ copies of the space transfer matrix $\mathbb{W}_t$ as
\be \label{eq:dualTracePowRho}
\tr\left[\rho_{A}^{n}(t)\right]= 
\tr\left[\big(\mathbb{W}_t^{\otimes n}\big)^{\abs*{\overline{A}}} 
\eta_{2n}^{\dagger}
\big(\!\!\left.\mathbb{W}_t^{\ast}\right.^{\otimes n}\!\!\big)^{\abs*{A}} 
\eta_{2n}\right],
\ee
where $\eta_{2n}$ represents a shift for one copy in the space of $2n$ replicas, and the complex conjugate of the transfer matrix comes from the exchange of conjugate and non-conjugate copies (cf.\ Fig.~\ref{fig:QCslope}). 

For $L-|A|$ and $|A|$ both larger than $t$ the powers of the transfer matrix can again be replaced by the fixed points. In this way we obtain 
\be \label{eq:trPowfps}
\lim_{\abs{A}\to\infty}
\lim_{L\to\infty}\tr\left[\rho_{A}^{n}(t)\right]=
\tr\left[(M_{{\rm L},t}^{\dagger}M_{{\rm R},t}^{\phantom{\dagger}})^n\right]^2,
\ee
which is schematically depicted on the r.h.s.\ of Fig.~\ref{fig:QCslope}. Thus, we find the following succinct expression for the slope~\eqref{eq:slope}
\be
\label{eq:slopeQC}
s_{\alpha}=\frac{1}{1-\alpha}\lim_{t \to \infty} 
\frac{1}{t}\ln\tr\left[(M_{{\rm L},t}^{\dagger}M^{\phantom{\dagger}}_{{\rm R},t})^\alpha\right],\quad \alpha\in\mathbb R,
\ee
where we analytically continued the matrix power. 

On the other hand, the stationary entropy density~\eqref{eq:density} is 
expected to coincide with the R\'enyi entropy of the reduced density
matrix of the stationary state, i.e.,
\be \label{eq:densityFin}
 d_{\alpha}=\frac{1}{1-\alpha}\lim_{\abs{A}\to\infty}\frac{1}{\abs{A}}
 \ln\tr\left[\rho_{{\rm st},A}^\alpha\right].
\ee
A comparison between the expressions~\eqref{eq:slopeQC} and~\eqref{eq:densityFin} shows that the slope can be written as a density of R\'enyi entropy as follows 
\be
s_{\alpha}= \frac{1}{1-\alpha}\lim_{t\to\infty}\frac{1}{t}
 \ln\tr\left[\tilde \rho_{{\rm st},t}^\alpha\right],
 \label{eq:slopetodensity}
\ee  
where we introduced the \emph{pseudo density matrix}
\be
\tilde{\rho}_{{\rm st},t}=M_{{\rm L},t}^{\dagger} M^{\phantom{\dagger}}_{{\rm R},t}.
\label{eq:rhotilde}
\ee
To gain physical intuition on the meaning of $\tilde{\rho}_{{\rm st},t}$ let us begin considering $M_{{\rm L},t}$ and $M_{{\rm R},t}$. These matrices are by definition the fixed points, or stationary states, of the space evolution (from left to right and right to left respectively). Moreover, as shown in Appendix~\ref{sec:fixedPointsCircuits}, they fulfil
\be
M_{{\rm R},t}=m_{{\rm R},t}^{\phantom{\dag}} m^{\dag}_{{\rm R},t}\,,\qquad\qquad M_{{\rm L},t}=m_{{\rm L},t}^{\phantom{\dag}} m^{\dag}_{{\rm L},t},
\label{eq:positivityMLMR}
\ee
implying that they are Hermitian and positive. This fact can be understood by recalling that, although not unitary, the evolution in space is a \emph{hybrid quantum evolution}~\cite{ippoliti2021postselectionfree}, i.e., it preserves positivity and Hermiticity.

Eq.~\eqref{eq:positivityMLMR} guarantees that $ \tilde{\rho}_{{\rm st},t}$ has real, non-negative eigenvalues, and, moreover, the normalisation condition \eqref{eq:fixedpointnormalization} gives  
\be
\tr \tilde{\rho}_{{\rm st},t}=1\,.
\ee 
Combining all the above facts together we see that, even though $\tilde{\rho}_{{\rm st},t}$ is not Hermitian and hence it cannot be interpreted as a proper quantum mechanical density matrix, it has many properties of reduced density matrices~\footnote{Note that \eqref{eq:slopetodensity} can be equivalently written in terms of the Hermitian pseudo density matrix $\tilde{\rho}_{t}= m^{\dag}_{{\rm L},t}m_{{\rm R},t}^{\phantom{\dag}} m^{\dag}_{{\rm R},t}m_{{\rm L},t}^{\phantom{\dag}}$. This matrix, however, is not stationary under the space evolution}. 

We note that a correspondence similar to \eqref{eq:slopetodensity} between entanglement slope in the original model and ``steady-state entanglement" in the dual model has been recently unveiled in Refs.~\cite{ippoliti2021fractal} and~\cite{ippoliti2021postselectionfree}. In particular, although considering a slightly different setting (they studied entanglement evolution in the presence of edge decoherence), these references expressed the entanglement dynamics in terms of states in the time lattice that in our language correspond to $\ket{M_{{\rm L},t}}$ and $\ket{M_{{\rm R},t}}$. Here, however, we provided two significant advances. First, we showed how to establish the correspondence when the evolution in the original model is purely unitary. Second, we provided a direct interpretation of $\ket{M_{{\rm L},t}}$ and $\ket{M_{{\rm R},t}}$ as fixed points of the space transfer matrix (or influence matrices). This, in turn, allows us to view $\tilde{\rho}_{{\rm st},t}$ as a stationary state of the space evolution.

In the special case when right and left fixed points coincide, $\tilde{\rho}_{{\rm st},t}$ is Hermitian, positive definite, normalised to one, and invariant under the space evolution. Therefore, it is a stationary density matrix of the space evolution. This happens, for instance, for \emph{dual-unitary quantum circuits}~\cite{bertini2019exact} evolving from compatible --- or solvable --- initial states~\cite{bertini2019entanglement,piroli2020exact}. Indeed, these systems and states are designed in such a way that the evolution in space is again a unitary brickwork quantum circuit, and, therefore, is equivalent to the time evolution. In particular, we have~\cite{bertini2019entanglement,piroli2020exact} 
\be
M_{{\rm R},t} = M_{{\rm L},t} = \frac{\1_{2t}}{d^{t}}, 
\ee
where $\1_x$ is the identity matrix acting on $x$ qudits. Therefore, in this case the pseudo density matrix coincides with the infinite temperature state, i.e.,
\be
\tilde{\rho}_{{\rm st},t} = \frac{\1_{2t}}{d^{2t}} \,.
\ee
Note that this is indeed the stationary state reached by the subsystem $[0,t]$ of the time lattice under space evolution. Thus, for dual-unitary circuits \eqref{eq:slopetodensity} is rewritten as 
\be
s_{\alpha}=\tilde d_{\alpha},
\ee
where $\tilde d_\alpha$ in \eqref{eq:slopetodensity} is the stationary density of entropy in the dual model. For dual-unitary circuits one can also repeat the above reasoning to compute the entanglement growth between a subsystem of the time lattice and the rest. This gives  
\be
\label{eq:sstilded}
\tilde s_{\alpha} = d_\alpha\,.
\ee

\section{Spacetime swap in relativistic quantum field theories}
\label{sec:qft}
Let us now change the setting and consider quantum field theories in 1+1 dimensions, i.e., generic quantum systems defined in a continuous spacetime. Since in these systems space and time are both continuous they are again closed under spacetime swap. Therefore, we expect that one can again establish a direct correspondence between slope and density of R\'enyi entropies. 

In 1+1 dimensional quantum field theories, however, it is more convenient to perform a slight variation of the spacetime swap. Specifically, instead of directly exchanging space and time here we consider the following analytic continuation
\be
(t,x)\mapsto(-i x,i t),
\label{eq:mirror}
\ee
which corresponds to an exchange of space and time in the Euclidean formulation of the theory~\cite{zamolodchikov1990thermodynamic}. In the string-theory literature the dual model obtained via the mapping \eqref{eq:mirror} is often referred to as the \emph{mirror model}~\cite{arutyunov2007on, arutyunov2009thermodynamic, tongeren2016introduction}.

Our key observation is that if one considers a 1+1 dimensional relativistic invariant quantum field theory, crossing symmetry implies that the mirror model coincides with the original one in the bulk. Therefore, in this setting 1+1 dimensional relativistic quantum field theories play a role similar to the dual-unitary circuits considered in the previous section. This means that, assuming appropriate compatible initial states, one should have
\be
s_\alpha= i \tilde d_\alpha,\qquad d_\alpha=- i \tilde s_\alpha\,,
\label{eq:correspondence}
\ee
where the quantities with the tilde denote the slope and density of R\'enyi entropy in the mirror model. To be more concrete and gain some intuition we begin by proving \eqref{eq:correspondence} for non-interacting, fermionic 1+1 dimensional quantum field theories. 

\subsection{Proof of (\ref{eq:correspondence}) for free theories}

Let us focus on a non-interacting quantum field theory of fermions in 1+1 dimensions. We consider the quench problem with the system initially prepared in the Gaussian state
\be
\ket{\Psi_0} = \exp\left(\int \frac{{\rm d}\mu}{2\pi}
K(\mu) \psi^\dag(-\mu)\psi^\dag(\mu)\right)\ket{0},
\label{eq:freeBS}
\ee 
where $\psi^\dag(\mu)$ is a creation operator for fermionic modes of rapidity $\mu$, $\ket{0}$ is the vacuum state for fermions, and $K(\mu)$ is an odd function such that 
\be
\int \frac{{\rm d}\mu}{2\pi} \abs{K(\mu)}^2<\infty\,.
\ee
Since a state of the form \eqref{eq:freeBS} produces pairs of correlated quasiparticles, the asymptotic slope \eqref{eq:slope} can be computed using the quasiparticle picture~\cite{calabrese2005evolution} 
\be
\!\!s_\alpha \!=\frac{2 m}{1-\alpha}\!\int_0^\infty\!\! \frac{{\rm d}\mu}{2\pi}\,\sinh\mu\, 
\ln\left[(1- \vartheta(\mu))^{\alpha} \!+\!  {\vartheta(\mu)^{\alpha}} \right].
\label{eq:slopeqp}
\ee
Here 
\be
\vartheta(\mu) = \frac{1}{1+\abs{K(\mu)}^2}= \vartheta(-\mu),
\label{eq:varthetaK}
\ee
is the \emph{occupation}, or \emph{filling} function, of the free mode with rapidity $\mu$ and we used the explicit form of the relativistic dispersion relation
\be
\varepsilon(\mu)=m \cosh \mu,\qquad p(\mu)=m \sinh\mu\,,
\ee
where $m$ is the mass of the fermions and we set the speed of light to one. Next, we recall that for a stationary state described by a filling function $\vartheta(\mu)$ the density of Renyi entropy reads as 
\be
\!d_\alpha \!=\frac{2 m}{1-\alpha}\!\int_0^\infty\! \frac{{\rm d}\mu}{2\pi}\,\cosh\mu\, \ln\left[(1- \vartheta(\mu))^{\alpha} \!+\!  {\vartheta(\mu)^{\alpha}} \right].\!
\label{eq:densityoriginalfree}
\ee
We then proceed by observing that at the level of rapidities the transformation
\eqref{eq:mirror} becomes~\cite{tongeren2016introduction}
\be
\mu \mapsto i \frac{\pi}{2}-\mu,
\label{eq:mirrormu}
\ee
therefore in the mirror model the occupation corresponding to $\vartheta(\mu)$ reads as
\be
\tilde \vartheta(\mu) = \vartheta(i\tfrac{\pi}{2}-\mu).
\label{eq:thetatilde}
\ee
Plugging this definition into the equation for the slope~\eqref{eq:slopeqp} we have  
\begin{align}
\!\!s_\alpha \!=&\frac{2m}{1-\alpha}\!\!\int_0^\infty\!\! \frac{{\rm d}\mu}{2\pi}\,\sinh\mu\, \ln\!\left[\!(1\!-\! \tilde\vartheta(i\tfrac{\pi}{2}\!-\!\mu))^{\alpha} \!\!+\!  {\tilde\vartheta(i\tfrac{\pi}{2}\!-\!\mu)^{\alpha}}\!\right] \notag\\
=&\frac{2i m}{1-\alpha}\int_{\Gamma}\frac{{\rm d}\mu}{2\pi}\,\cosh\mu\, \ln\left[(1- \tilde\vartheta(\mu))^{\alpha} \!+\!  {\tilde\vartheta(\mu)^{\alpha}} \right]\!,
\label{eq:mirrorslope}
\end{align}
where $\Gamma$ is the positively oriented contour parametrised by
$i\tfrac{\pi}{2}-x$ with $x \leq 0$. Comparing~\eqref{eq:mirrorslope} with
\eqref{eq:densityoriginalfree} we see that the former can be interpreted as the
density of entropy in the mirror model times $i$. We note that in the mirror model
the integration is performed over $\Gamma$ to keep $\tilde\vartheta(x)$ well defined
and rapidly decaying.

Analogously, substituting the definition~\eqref{eq:thetatilde} into the expression for the density~\eqref{eq:densityoriginalfree} we see that the density of entropy in the original model can be interpreted as $-i$ times the slope of the mirror model
\be
\!d_\alpha \!=\frac{-2i m}{1-\alpha}\!\int_{\Gamma}\! \frac{{\rm d}\mu}{2\pi}\,\sinh\mu\, \ln\left[(1- \tilde \vartheta(\mu))^{\alpha} \!+\!  {\tilde \vartheta(\mu)^{\alpha}} \right].\!
\label{eq:mirrordensity}
\ee 

\subsection{Slope of R\'enyi entropies in interacting integrable theories}
\label{sec:iqft}

Our next step is to use the correspondence \eqref{eq:correspondence} to find a prediction for the slope in the presence of interactions. To this end, a convenient setting to consider is that of massive integrable quantum field theories with a single species of excitations. We consider integrable quantum systems because for these systems the stationary density can be computed exactly using the \emph{quench action} approach~\cite{alba2017quench}, while we focus on theories with a single species of particle excitations for simplicity. Note that for this family of integrable quantum field theories, a similar argument based on the mapping~\eqref{eq:mirror} has been employed in Ref.~\cite{castroalvaredo2016emergent} to compute the expectation values of currents in stationary states.
 
In integrable quantum field theories the scattering is elastic and completely factorised. Therefore, it is fully determined by the two-particle scattering matrix $ S(\mu)$. This is a meromorphic function of the rapidity in the \emph{physical strip} $\mathcal S= \{0 \leq {\rm Im}\,\mu\leq \pi\}$, fulfilling \emph{unitarity},
\emph{crossing symmetry}, and \emph{real analyticity} 
\begin{align}
&S(\mu)S(-\mu)=1,  & &\mu\in\mathcal S,
\label{eq:unitarity}\\
&S(\mu)=S(i\pi-\mu), & &\mu\in\mathcal S,
\label{eq:crossing}\\
&S(\mu)^*=S(-\mu), & &\mu\in \mathbb R,
\label{eq:realanalyticity}\\
&S(i \mu)^*=S(i\mu), & &\mu\in [0,\pi]. 
\label{eq:realanalyticity2}
\end{align}
A powerful method to describe the thermodynamics of interacting integrable theories is provided by the \emph{thermodynamic Bethe ansatz} (TBA)~\cite{zamolodchikov1990thermodynamic}. In essence, with this method one describes stationary macrostates specifying the density of particle excitations that they contain. This is possible because in these systems particle excitations are stable and hence their densities are conserved. 

In particular, we can again describe macrostates using the filling
function $\vartheta(\mu)$, which describes the fraction of available states
that are occupied by the particles. Now, however, the density of available
states, denoted by $\rho_t(\mu)$, is \emph{not} a simple Jacobian as in the
non-interacting case. Because of the interactions it depends on the filling function through
the following integral equation~\cite{zamolodchikov1990thermodynamic}
\be
\rho_t(\mu)=\frac{m}{2\pi}\cosh(\mu)+
\!\!\int\frac{{\rm d}\mu'}{2\pi}T(\mu-\mu')\rho_t(\mu') \vartheta(\mu').
\label{thermo_bethe_eq}
\ee
Here we introduced the \emph{scattering kernel} $T(\mu)$, which is given by the
logarithmic derivative of the scattering matrix 
\be
T(\mu)= -i \frac{{\rm d}}{{\rm d}\mu}\ln S(\mu)\,.
\ee
Since we are considering theories evolving from non-equilibrium initial states, the mirror model will have a nontrivial boundary in space. Here we are interested in the case where this boundary does not break the integrability of the theory, so that we can use the result of Ref.~\cite{alba2017quench} for the density of R\'enyi entropy, therefore, we have to consider initial states generating integrable boundary conditions. These states are well known in the context of quantum integrability, and they are typically referred to as integrable boundary states~\cite{ghoshal1994boundary}. Note that integrable boundary states need regularisation to be considered as initial states of quench problems~\cite{fioretto2010quantum,bertini2014quantum, bertini2016quantum, sotiriadis2014boundary, horvath2016initial}. This is essentially due to the fact that they give non-zero weight to configurations involving particles with infinite energy.

A regularised integrable boundary state can be thought of as the generalisation of \eqref{eq:freeBS}. It is obtained by replacing $\psi^\dag(\mu)$ with the operators creating the
stable excitations~\cite{zamolodchikov1979factorised} and requiring $K(\mu)$
to satisfy
\be
K(\mu)=S(2\mu) K(-\mu),
\label{eq:boundaryreflection}
\ee
instead of being odd. The filling function corresponding to such a state is then obtained as the
solution to the following integral equation~\cite{fioretto2010quantum,bertini2016quantum}  
\be
\!\!\!\ln\!\frac{\vartheta(\mu)}{1\!-\!\vartheta(\mu)}\!=\!\ln{|K(\mu)|^{2}}\!-\!\!\int\!\!\frac{{\rm d}\mu'}{2\pi}T(\mu^\prime-\mu)\ln\!\left[1\!-\!\vartheta(\mu')\right].\!\!
\label{eq:saddle_point}
\ee
Note that \eqref{eq:realanalyticity} and \eqref{eq:boundaryreflection} imply
that $\vartheta(\mu)$ is even. 

Having introduced the necessary formalism we are finally in a position to write
an expression for $s_\alpha$. Our starting point is the exact expression for
$d_\alpha$ in interacting integrable models derived in
Ref.~\cite{alba2017quench}. Specialising it to the case at hand we can express
it as
\be
\!\!\!\!d_\alpha \!=\frac{2m}{1-\alpha}\!\int_0^\infty\!\! 
\frac{{\rm d}\mu}{2\pi}\,\cosh\mu\, \ln\!\!\left[(1- \vartheta(\mu))^{\alpha} \!+\!  
\frac{\vartheta(\mu)^{\alpha}}{x_\alpha(\mu)} \right],
\label{eq:densityiQFT}
\ee
where the auxiliary function $x_\alpha(\mu)$ is the solution to the following
integral equation 
\be
\begin{aligned}
\ln x_\alpha(\lambda) = & \!\int_0^\infty\!\!\! {{\rm d} \mu} \,\, 
    (T( \mu- \lambda)+T(\mu+\lambda)) \label{eq:xTBARel}\\
&\qquad\times\ln\left[(1- \vartheta(\mu))^{\alpha} \!\!+  
    \frac{\vartheta(\mu)^{\alpha}}{x_\alpha(\mu)} \right].
\end{aligned}
\ee
Proceeding as in the free case by substituting the definition \eqref{eq:thetatilde} of $\tilde \vartheta(\mu)$ in \eqref{eq:densityiQFT} we find 
\be
\!\!d_\alpha \!=\frac{-2 i m}{1-\alpha}\!\int_\Gamma \frac{{\rm d}\mu}{2\pi}\,\sinh\mu\, \ln\!\!\left[(1- \tilde \vartheta(\mu))^{\alpha} \!+\!  \frac{\tilde \vartheta(\mu)^{\alpha}}{\tilde y_\alpha(\mu)} \right]\!,
\ee
where we defined
\be
\tilde y_\alpha(\lambda)= x_\alpha(i\tfrac{\pi}{2}-\lambda).
\ee
The latter fulfils the following integral equation 
\be
\begin{aligned}
\ln \tilde y_\alpha(\lambda) = &\!\int_\Gamma\! {{\rm d} \mu} \,\, 
    (T( \mu- \lambda)-T(\mu+\lambda))\\
&\qquad\times\ln\left[(1- \tilde\vartheta(\mu))^{\alpha} \!\!+  
    \frac{\tilde \vartheta(\mu)^{\alpha}}{\tilde y_\alpha(\mu)} \right].
\end{aligned}
\ee
Where we used that, because of the crossing symmetry~\eqref{eq:crossing}
of the scattering matrix, the kernel satisfies the following relation,
\be
T(i\pi+\mu) = - T(\mu)\,.
\ee
Using then the correspondence \eqref{eq:correspondence} and rewriting
everything for the original model we finally find
\be
\!\!s_\alpha \!=\frac{2m}{1-\alpha}\!\int_0^\infty\!\! \frac{{\rm d}\mu}{2\pi}\,\sinh\mu\, 
\ln\!\!\left[(1-  \vartheta(\mu))^{\alpha} \!+\!  
\frac{ \vartheta(\mu)^{\alpha}}{ y_\alpha(\mu)} \right]
\label{eq:slopeiQFT}
\ee
with 
\begin{align}
\ln y_\alpha(\lambda) = &\!\int_0^\infty\! {{\rm d} \mu} \,\, (T( \mu- \lambda)-T(\mu+\lambda))\\
&\qquad\times\ln\left[(1- \vartheta(\mu))^{\alpha} \!\!+  
    \frac{\vartheta(\mu)^{\alpha}}{y_\alpha(\mu)} \right].\notag
\end{align}
This concludes our derivation of $s_\alpha$ for interacting integrable quantum
field theories with diagonal scattering. 

\section{Slope of R\'enyi entropies in generic TBA-integrable models}
\label{sec:TBA}

The argument leading to Eq.~\eqref{eq:slopeiQFT} can be applied to a much
larger class of TBA-solvable models. An immediate generalisation is obtained by
considering integrable quantum field theories with non-diagonal scattering as the sine-Gordon field theory. Indeed, since integrable boundary states also exist for these systems~\cite{ghoshal1994boundary}, one can directly repeat the treatment of the previous section. 

In fact, the existence of integrable boundary states is not limited to field
theories. Also in algebraic-Bethe-ansatz-integrable lattice systems there exist
initial states for which the system obtained by exchanging space and time is
integrable~\cite{pozsgay2013dynamical,piroli2017what,piroli2019integrableI,piroli2019integrableII,pozsgay2019integrable}.
This applies most directly to integrable systems with a discrete time
evolution~\cite{vanicat2018integrable}. In these systems the time evolution is
generated by an integrable transfer matrix and, by taking appropriate initial
states~\cite{pozsgay2013dynamical,piroli2017what,piroli2019integrableI,piroli2019integrableII,pozsgay2019integrable}, one
can ensure integrability of the (boundary) transfer matrix in
space~\cite{sklyanin1988boundary}. The case of lattice systems with continuous
time evolution can then be recovered by taking the \emph{Trotter
limit}~\cite{piroli2017quantum,piroli2018non},
i.e., sending the discrete time-step $\Delta t$ to zero, while keeping fixed the
real time $t = N\cdot \Delta t$ with $N$ being the number of steps.    

In light of these facts here we argue that Eq.~\eqref{eq:slopeiQFT} can be extended to all TBA-integrable systems by a simple generalisation of the TBA description. In particular we have to account for the following modifications.
\begin{enumerate}[label=(\roman*)]
    \item\label{mod1} Generic integrable models feature multiple species of
        quasiparticles~\cite{takahashi1999thermodynamics,zamolodchikov1990thermodynamic}.
        This means that in general quasiparticles are no longer specified only by their
        rapidity $\lambda\in \mathbb R$ but one also needs to introduce a
        discrete species index $n\in \mathbb N$. Effectively, this means
        that we have to make the replacement
        \be
        \lambda \mapsto (\lambda,n), 
        \ee
        in the arguments of all functions. Naturally, this also means that when
        integrating over the rapidity, also the sum over all the possible
        particle species has to be performed
        \be
        \int \!\!{{\rm d}\mu}\, f(\mu) \mapsto \sum_m\,
        \int \!\!{{\rm d}\mu}\, f_m(\mu),
        \ee
        where we followed the standard convention of reporting the species index in the subscript. Note also that the integration and summation boundaries depend on the specific model.
        Finally, to describe scattering among particles of different species, the scattering
        kernel needs to be generalised,
        \be
        T( \lambda- \mu) \mapsto 
        T_{nm}(\lambda,\mu).  
        \ee

        To keep track of rapidity and species index we employ the following
        compact notation
        \be
        \begin{aligned}
            (\lambda,n) &\equiv \boldsymbol \lambda, \\
            \sum_m\,\, \int \!\!{{\rm d}\mu}\, f_m(\mu)  
            &\equiv \int\!\! {{\rm d}\boldsymbol \mu}\, f(\boldsymbol \mu),\\
            T_{nm}( \lambda, \mu)  &\equiv T(\boldsymbol \lambda,\boldsymbol \mu).
        \end{aligned}
        \label{eq:shorthand}
        \ee

    \item\label{mod2}In generic TBA integrable systems the dispersion relation does not necessarily coincide with the relativistic one, therefore we make the replacement 
            \be
            m\cosh \lambda \mapsto \varepsilon(\boldsymbol \lambda),\qquad
            m\sinh \lambda \mapsto p(\boldsymbol \lambda)\,.
            \ee
            Here the parametrisation is chosen such that 
            $\varepsilon'(\boldsymbol{\lambda})=\varepsilon^{\prime}_{n}(\lambda)$,
            is always positive for $\lambda>0$. 
\end{enumerate}

Taking into account the modifications \ref{mod1} and \ref{mod2},
Eq.~\eqref{eq:slopeiQFT} is rewritten as 
\be
    \!\!\!s_{\alpha} \! = \frac{2}{1-\alpha} \!\int_+ \!\!\! {{\rm d}\boldsymbol \mu}\, 
    \frac{\varepsilon'(\boldsymbol \mu)}{2\pi}\,  
    \ln\left[(1- \vartheta(\boldsymbol \mu))^{\alpha} \!\!+  
    \frac{\vartheta(\boldsymbol \mu)^{\alpha}}{y_\alpha(\boldsymbol \mu)} \right]\!,
    \label{eq:conjecture}
\ee
where we introduced the auxiliary function
\be
\begin{aligned}
    \ln y_\alpha(\boldsymbol \lambda) = 
    & \!\int_+ \!\!\! {{\rm d}\boldsymbol \mu} \,\, 
    (T(\boldsymbol \mu,\boldsymbol \lambda)-T(\boldsymbol \mu,-\boldsymbol \lambda)) 
    \label{eq:yTBA}\\
    &\qquad\times\ln\left[(1- \vartheta(\boldsymbol \mu))^{\alpha} \!\!+  
    \frac{\vartheta(\boldsymbol \mu)^{\alpha}}{y_\alpha(\boldsymbol \mu)} \right]\!.
\end{aligned}
\ee
Here $(\cdot)'$ denotes a derivative with respect to the real rapidity $\mu$
and the subscript $+$ indicates that the integral range is restricted to positive
rapidities. To express~\eqref{eq:conjecture} we implicitly used that,
apart from fine tuned cases, integrable initial states produce reflection symmetric rapidity
distributions~\cite{ghoshal1994boundary,piroli2017what,piroli2019integrableI,piroli2019integrableII,pozsgay2019integrable}.

Once again \eqref{eq:conjecture} closely parallels the expression for the
density of R\'enyi entropy in the post quench stationary state described by the
rapidity distribution $\vartheta(\boldsymbol \mu)$. Indeed, for a
reflection-symmetric $\vartheta(\boldsymbol \mu)$ we have~\cite{alba2017quench} 
\be
\!d_{\alpha} \! = \frac{2}{1-\alpha}\!\int_+ \!\!\! 
{{\rm d}\boldsymbol \mu}\, \frac{|p'(\boldsymbol \mu)|}{2\pi}\, 
\ln\left[(1- \vartheta(\boldsymbol \mu))^{\alpha} \!\!+  
\frac{\vartheta(\boldsymbol \mu)^{\alpha}}{x_\alpha(\boldsymbol \mu)} \right]\!,\!
\label{eq:densityTBA}
\ee
with
\be\label{eq:xTBA}
\begin{aligned}
    \ln x_\alpha(\boldsymbol \lambda) = & \!\int_+ 
    \!\!\! {{\rm d}\boldsymbol \mu} \,\, 
    (T(\boldsymbol \mu,\boldsymbol \lambda)+T(\boldsymbol \mu,-\boldsymbol \lambda)) \\
    &\qquad\times\ln\left[(1- \vartheta(\boldsymbol \mu))^{\alpha} \!\!+ 
    \frac{\vartheta(\boldsymbol \mu)^{\alpha}}{x_\alpha(\boldsymbol \mu)} \right].
\end{aligned}
\ee
In what follows we present strong evidence for the validity of~\eqref{eq:conjecture} by providing four nontrivial consistency checks. In particular, in Sec.~\ref{sec:ff} we show that Eq.~\eqref{eq:conjecture} reduces to the exact free-fermion result~\cite{alba2017quench} when the interaction kernel vanishes. Next, in Sec.~\ref{sec:vonNeumann} we prove that in the limit $\alpha\to 1$ the expression recovers the quasiparticle prediction~\cite{alba2017entanglement} for the slope of the von Neumann entanglement entropy. In Sec.~\ref{sec:Rule54} we show that~\eqref{eq:conjecture} agrees with the exact result of Refs.~\cite{klobas2021exact,klobas2021entanglement} for a specific interacting integrable model treatable by TBA, i.e., the cellular automaton \emph{Rule 54}~\cite{bobenko1993two}. Finally, in Sec.~\ref{sec:XXZ} we compare Eq.~\eqref{eq:conjecture} with exact numerical results for the XXZ spin-1/2 chain.

Before that, however, we rewrite the expression~\eqref{eq:conjecture} in an equivalent form, which is more convenient for parts of the upcoming analysis. To this end, we introduce two quantities that are very convenient in the TBA analysis of integrable systems, namely \emph{total density} $\rho_t(\boldsymbol \lambda)$ and \emph{dressed velocity} $v(\boldsymbol \lambda)$. The former is the direct generalisation of the density of available states introduced in Sec.~\ref{sec:iqft} and is defined as the solution to the following integral equation  
\be
    \rho_{t}(\boldsymbol \lambda)  = \frac{|p'(\boldsymbol \lambda)|}{2\pi} 
    - \!\!\int\!\! {{\rm d}\boldsymbol \mu}\, T(\boldsymbol \lambda,\boldsymbol \mu)  
    \vartheta(\boldsymbol \mu) \rho_{t}(\boldsymbol \mu) \,.
\label{eq:totalrho}
\ee
The latter is the velocity of quasiparticle excitations in the state described by $\vartheta(\boldsymbol \mu)$ and is determined by
\be
    \!\!v(\boldsymbol \lambda) \rho_{t}(\boldsymbol \lambda)  = 
    \frac{\varepsilon'(\boldsymbol \lambda)}{2\pi} - \!\!\int\!\! {{\rm d}\boldsymbol \mu}\, 
    T(\boldsymbol \lambda,\boldsymbol \mu)  \vartheta(\boldsymbol \mu) v(\boldsymbol \mu) 
    \rho_{t}(\boldsymbol \mu)\,.\!\!
\label{eq:vtotalrho}
\ee
This quantity plays a crucial role in the quench dynamics, which was first observed in Ref.~\cite{bonnes2014light}.

Inserting Eqs.~\eqref{eq:totalrho} and \eqref{eq:vtotalrho} into Eqs.~\eqref{eq:densityTBA}, and~\eqref{eq:conjecture},
we obtain the following equivalent expressions for density
\begin{align}
&\begin{aligned}
    d_\alpha &= \frac{2}{1-\alpha} \!\int_+ \!\!\! 
    {{\rm d}\boldsymbol \mu}\, \rho_{t}(\boldsymbol \mu) 
    \ln\!\left[\!(1- \vartheta(\boldsymbol \mu))^{\alpha} \!\!+  
    \frac{\vartheta(\boldsymbol \mu)^{\alpha}}{x_\alpha(\boldsymbol \mu)} \right]\\
    &+ \frac{2}{1-\alpha} \!\int_+ \!\!\! {{\rm d}\boldsymbol \mu} \, \rho(\boldsymbol \mu)\, 
    {\ln x_{\alpha}(\boldsymbol \mu)},
    \label{eq:convenientdensity}
\end{aligned}
\end{align}
and slope
\begin{align}
&\begin{aligned}
    \!s_\alpha \! &=\! \frac{2}{1-\alpha}  \!\int_+ \!\!\! {{\rm d}\boldsymbol \mu}\, 
    \rho_{t}(\boldsymbol \mu) v(\boldsymbol \mu)
    \ln\!\left[\!(1- \vartheta(\boldsymbol \mu))^{\alpha} \!\!+  
    \frac{\vartheta(\boldsymbol \mu)^{\alpha}}{y_\alpha(\boldsymbol \mu)} \right]\!\\
    &+\!\frac{2}{1-\alpha} \!\int_+ \!\!\! {{\rm d}\boldsymbol \mu} \, 
    \rho(\boldsymbol \mu)v(\boldsymbol \mu){\ln y_{\alpha}(\boldsymbol \mu) }.
    \label{eq:convenientslope}
\end{aligned}
\end{align}

\subsection{Free fermions}\label{sec:ff}
Our first check concerns free-fermionic systems. In this case, the absence of 
interactions permits \emph{ab initio}
calculations~\cite{calabrese2005evolution,fagotti2008dinamica,fagotti2008evolution,castroalvaredo2019entanglement},
which prove the validity of quasiparticle picture also for R\'enyi-$\alpha$ entropies.
Therefore we expect to recover the quasiparticle result in the limit of the vanishing
interaction kernel.

Indeed, for $T({\boldsymbol \lambda},{\boldsymbol \mu})=0$, Eq.~\eqref{eq:yTBA} gives  
\be
y_{\alpha}(\mu)=1,\qquad \forall \mu\,,
\ee
and thus~\eqref{eq:conjecture} becomes 
\be
\!\!\!s_{\alpha} \! = \frac{2}{1-\alpha} \!\int_+ \!\!\! {{\rm d}\boldsymbol \mu}\, \frac{\varepsilon'(\boldsymbol \mu)}{2\pi}\,  \ln\left[(1- \vartheta(\boldsymbol \mu))^{\alpha} \!\!+  {\vartheta(\boldsymbol \mu)^{\alpha}} \right]\!,
\ee
which is precisely the quasiparticle-picture result~\cite{alba2017quench}.

\subsection{Von Neumann}\label{sec:vonNeumann}
To recover the prediction for the von-Neumann entanglement entropy we
consider the limit $\alpha\to1$ of the expressions~\eqref{eq:convenientslope},
and~\eqref{eq:yTBA}. We begin by noting that
\be
y_1(\mu)=1, 
\ee
which follows from the observation that in the $\alpha\to 1$ limit the function
$y_{1}(\mu)=1$ solves Eq.~\eqref{eq:yTBA}, combined with the standard TBA
assumption of uniqueness of solutions. Evaluating the limit $\alpha\to 1$ in
Eq.~\eqref{eq:convenientslope} 
we then have
\begin{align}
    \lim_{\alpha\to1} s_\alpha =& 2 \!\int_+ \!\!\! {{\rm d}\boldsymbol \mu}\, v(\boldsymbol \mu)
    s(\boldsymbol \mu).
    \label{eq:vonNeumann}
\end{align}
Here we used that the terms containing 
$\partial_\alpha y_\alpha(\mu)$
cancel, and introduced 
$s({\boldsymbol \mu})$ for the density of the Yang-Yang entropy,
\be \label{eq:YYentropydensity}
\mkern-16mu s(\boldsymbol \mu)\!=\!- \rho_{t}(\boldsymbol \mu)
\!\left[(1\!-\!\vartheta(\boldsymbol \mu))\!\ln(1\!- \!\vartheta(\boldsymbol \mu))\!
+\! \vartheta(\boldsymbol \mu) \ln\vartheta(\boldsymbol \mu)\right]\!.\mkern-16mu
\ee
As promised, \eqref{eq:vonNeumann} recovers the quasiparticle
prediction~\cite{alba2017entanglement} for the slope of the von Neumann entanglement entropy.

\subsection{Rule 54} \label{sec:Rule54}
\begin{figure}
    \includegraphics[width=\columnwidth]{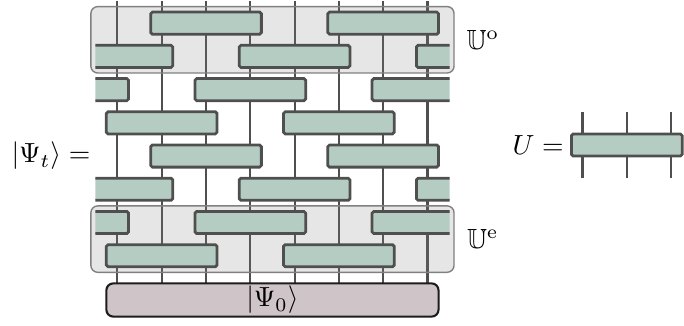}
    \caption{\label{fig:Rule54} Schematic representation of time evolution of
    Rule 54. In each time-step the three-site gates $U$ are applied either to 
    even ($\mathbb{U}^{\rm e}_{L}$) or odd ($\mathbb{U}^{\rm o}_L$) triplets of
    sites. Note that gates that overlap on at most one site commute.
    }
\end{figure}
Our next consistency check involves what is so far the only known exact result for the slope of R\'enyi entropies in an interacting integrable system: the result of Refs.~\cite{klobas2021exact,klobas2021entanglement} for the Rule 54 cellular automaton. The model was introduced in Ref.~\cite{bobenko1993two}, and has recently been identified as one of the simplest examples of interacting integrable systems~\cite{buca2021rule}, allowing for the exact description of many non-equilibrium properties, both in the classical~\cite{prosen2016integrability,prosen2017exact,inoue2018two,buca2019exact,klobas2019time,klobas2020matrix,klobas2020space}, and the quantum realm~\cite{gopalakrishnan2018facilitated,gopalakrishnan2018operator,gopalakrishnan2018hydrodynamics,friedman2019integrable,alba2019operator,alba2021diffusion,klobas2021exact,klobas2021exactrelaxation,klobas2021entanglement,lopezpiqueres2022integrability}.

Rule 54 can be understood as a quantum circuit consisting of $3$-site local \emph{deterministic}
gates $U$ with the following matrix elements,
\be
U_{s_1 s_2 s_3}^{s_1^{\prime} s_2^{\prime} s_3^{\prime}}=
\delta_{s_1^{\prime},s_1}
\delta_{s_2^{\prime},\chi(s_1,s_2,s_3)}
    \delta_{s_3^{\prime},s_3},
\ee
where we introduced the binary function $\chi:\mathbb{Z}_2^{\times 3} \to \mathbb{Z}_2$
\be
    \chi(s_1,s_2,s_3)\equiv s_1+s_2+s_3+s_1 s_3\pmod{2}.
\ee
Time-evolution is given in two distinct time-steps
\be 
\mathbb{U}_{L}=\mathbb{U}^{\rm o}_L \mathbb{U}^{\rm e}_L,
\ee
which involve the gates $U$ applied at odd or even triplets of sites
(see Fig.~\ref{fig:Rule54} for an illustration)
\be
\mathbb{U}^{\rm e}_L=\prod_{j=0}^{L}
\Pi_{2L}^{-2j} 
U
\Pi_{2L}^{2j},\qquad
\mathbb{U}^{\rm o}_L=\Pi_{2L}^{-1} \mathbb{U}^{\rm e}\Pi_{2L}.
\ee
We recall that $\Pi_{2L}$ is a periodic shift for one site on the chain of $2L$ sites. The gate $U$ deterministically changes only the middle site depending on the state of the sites on both edges. Therefore all the local operators applied at the same step commute, and the dynamics is indeed one of a local quantum circuit, albeit with a slightly nonstandard geometry (cf. Fig.~\ref{fig:Rule54}). This allows us to use the ideas from Sec.~\ref{sec:duality} to formally express the slope $s_{\alpha}$. Moreover, as demonstrated in Refs.~\cite{klobas2021exact,klobas2021entanglement}, the solvability of the model allows for an exact calculation of the fixed points (cf. Sec.~\ref{sec:duality}), and hence of the entanglement slope, for a family of solvable initial states.

More concretely, for a quench from the state
\be 
\label{eq:initialStateCompBasis}
\ket{\Psi_0}=
\left(
\begin{bmatrix}
    1 \\ 0
\end{bmatrix} \otimes
\begin{bmatrix}
    \sqrt{1-\vartheta}\\
    \mathrm{e}^{i \varphi} \sqrt{\vartheta}
\end{bmatrix}
\right)^{\otimes L}
\ee
with $\varphi\in[0,2\pi]$ and $\vartheta\in[0,1]$, the asymptotic slope of
the R\'enyi-$\alpha$ entropy reads as  
\be
s_\alpha = \frac{2}{1-\alpha}
\ln\left[(1- \vartheta)^{\alpha} +  \frac{\vartheta^{\alpha}}{y_\alpha}\right],
\label{eq:exactresultR54}
\ee
where $y_\alpha$ is the only real and positive solution to the following equation 
\begin{align}
    \ln y = 2 \ln\left[(1- \vartheta)^{\alpha} +  \frac{\vartheta^{\alpha}}{y_\alpha}\right]. 
    \label{eq:exactresultyR54}
\end{align}

To compare the above exact expression with Eq.~\eqref{eq:conjecture} we need to
recall some facts about the TBA description of
the model~\cite{friedman2019integrable,buca2021rule}, and a quench from the
state~\eqref{eq:initialStateCompBasis}~\cite{klobas2021entanglement}.
\begin{enumerate}[label=(\roman*)]
    \item The TBA description of states relevant for this quench problem
        involve only two species of particles, left and right movers, labelled
        by $n=\pm 1$.
    \item The derivative of the dispersion relation $p_{n}^{\prime}(\lambda)$,
        $\varepsilon_n^{\prime}(\lambda)$, and the scattering kernel
        $T_{n m}(\lambda,\mu)$ are independent of the rapidities
        $\lambda,\mu\in[-\pi,\pi]$ and read as 
        \be
        \label{eq:TBArule54}
        \begin{gathered}
            p_n^{\prime}(\lambda)=n,\qquad \varepsilon_n^{\prime}(\lambda)=1,\\
            T_{n m}(\lambda,\mu)=\frac{n m}{2 \pi}.
        \end{gathered}
        \ee
    \item The stationary state that the system approaches after the quench
        from~\eqref{eq:initialStateCompBasis} corresponds to filling functions
        equal to the parameter $\vartheta$ of the initial state
        \be\label{eq:thetaQuenchRule54}
        \vartheta_{n}(\lambda)=\vartheta.
        \ee
\end{enumerate}

Rewriting Eqs.~\eqref{eq:conjecture} and~\eqref{eq:yTBA} with these properties in mind,
we find precisely Eqs.~\eqref{eq:exactresultR54} and~\eqref{eq:exactresultyR54}.

\subsection{XXZ model}
\label{sec:XXZ}
\begin{figure*}
    \centering
    (a)%
    \raisebox{-\totalheight+\baselineskip}[0pt][\totalheight]{\includegraphics[width=0.95\columnwidth]{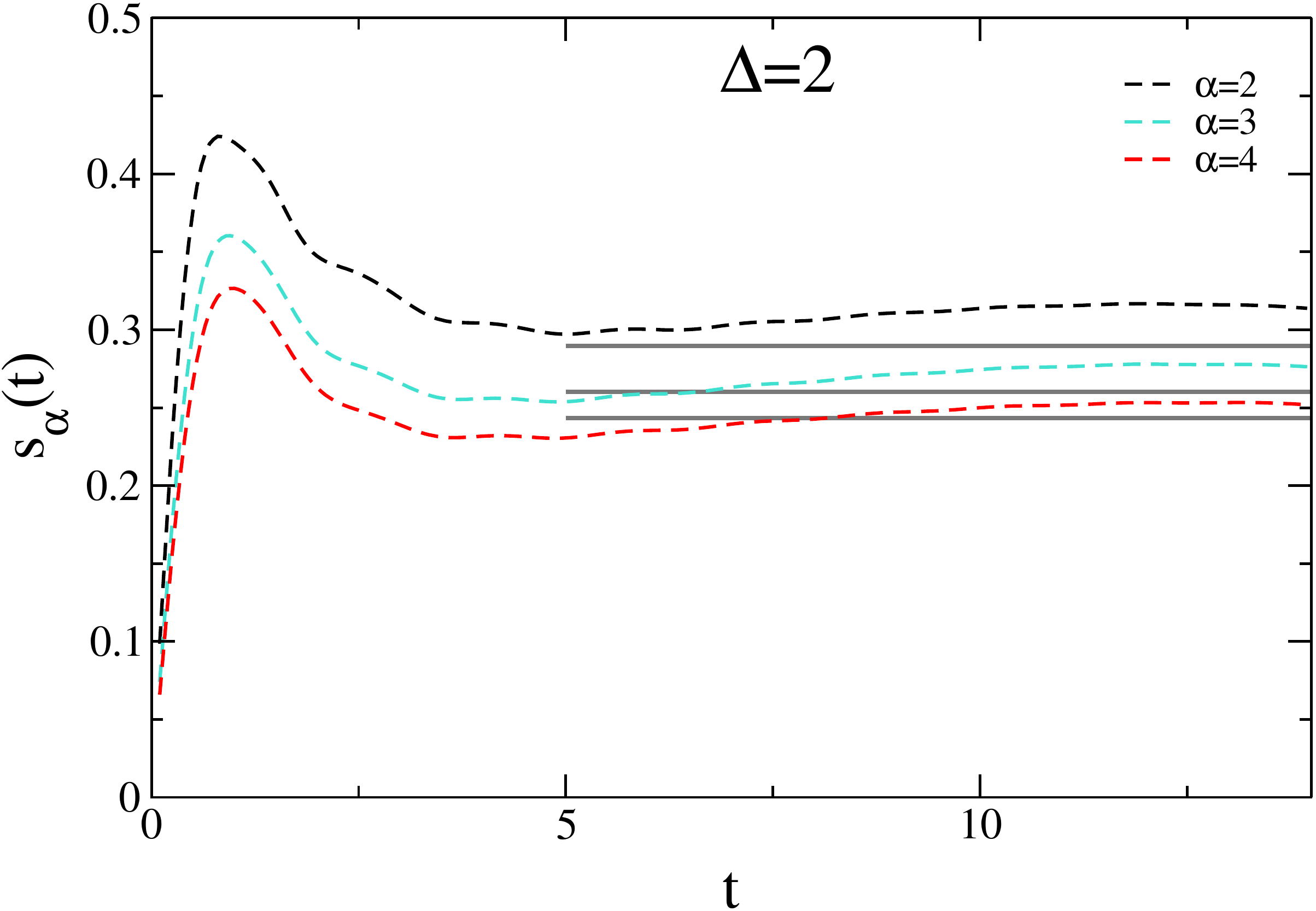}}\hfill
    (b)%
    \raisebox{-\totalheight+\baselineskip}[0pt][\totalheight]{\includegraphics[width=0.95\columnwidth]{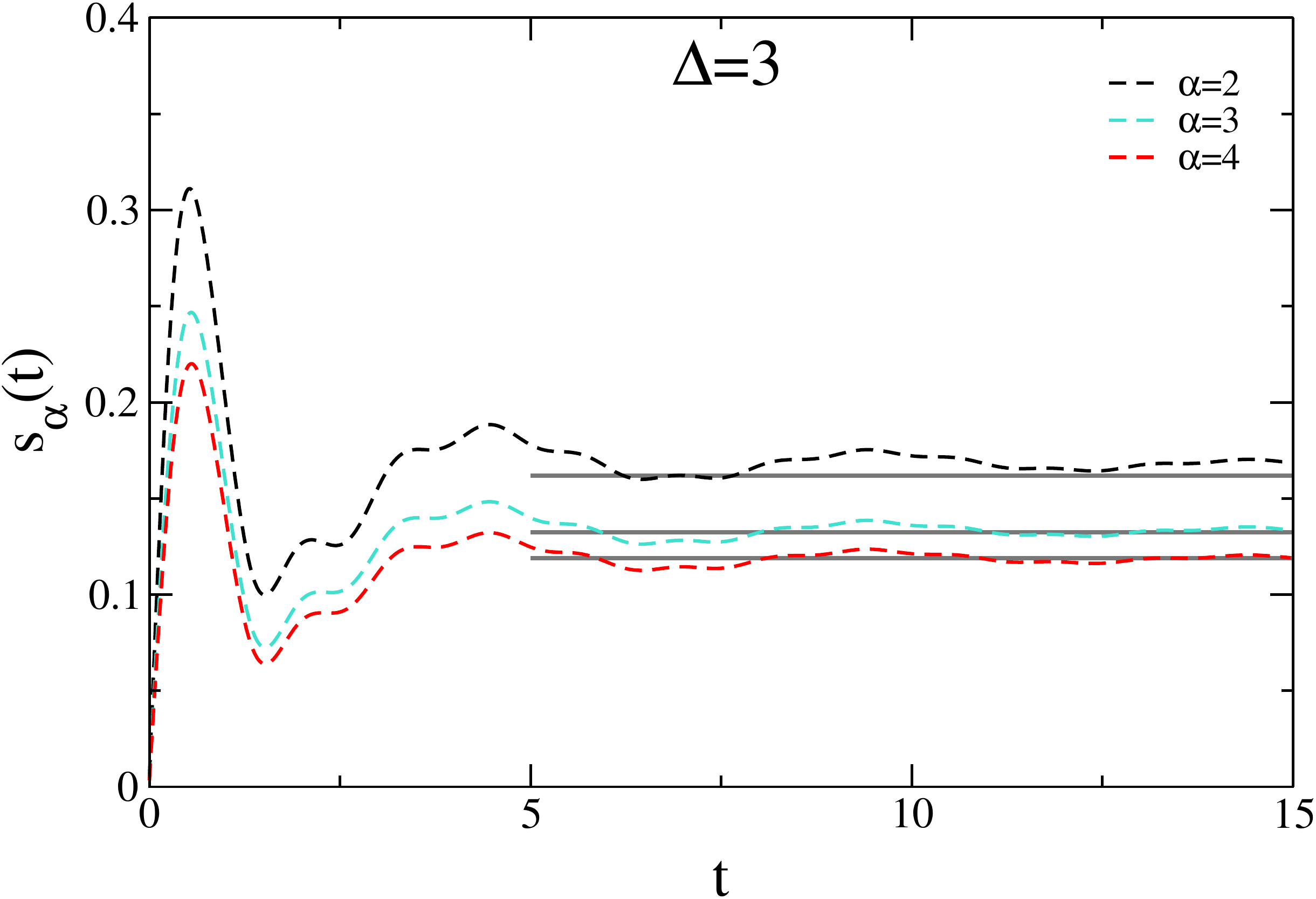}}\\[1em]
    (c)%
    \raisebox{-\totalheight+\baselineskip}[0pt][\totalheight]{\includegraphics[width=0.95\columnwidth]{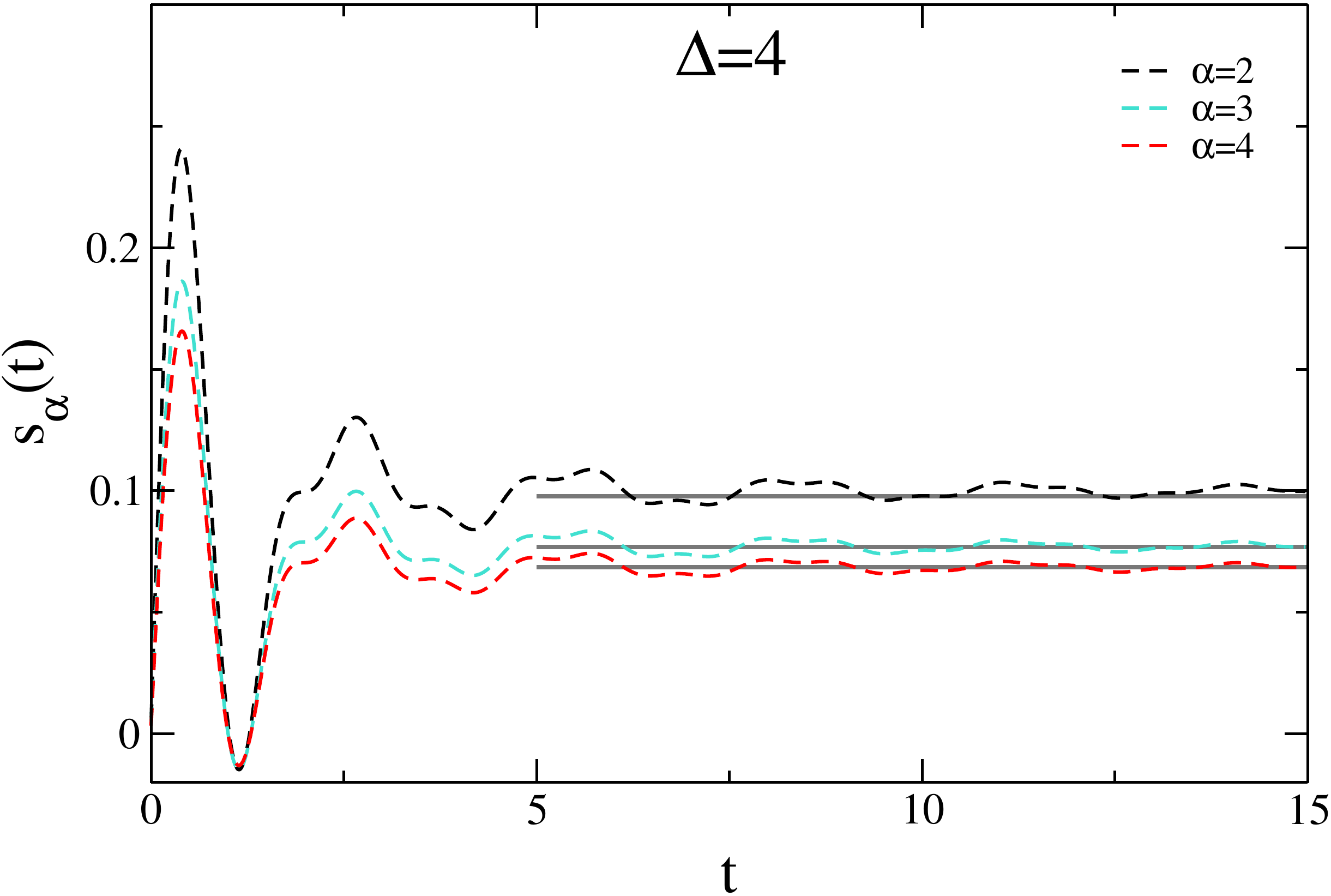}}\hfill
    (d)%
    \raisebox{-\totalheight+\baselineskip}[0pt][\totalheight]{\includegraphics[width=0.95\columnwidth]{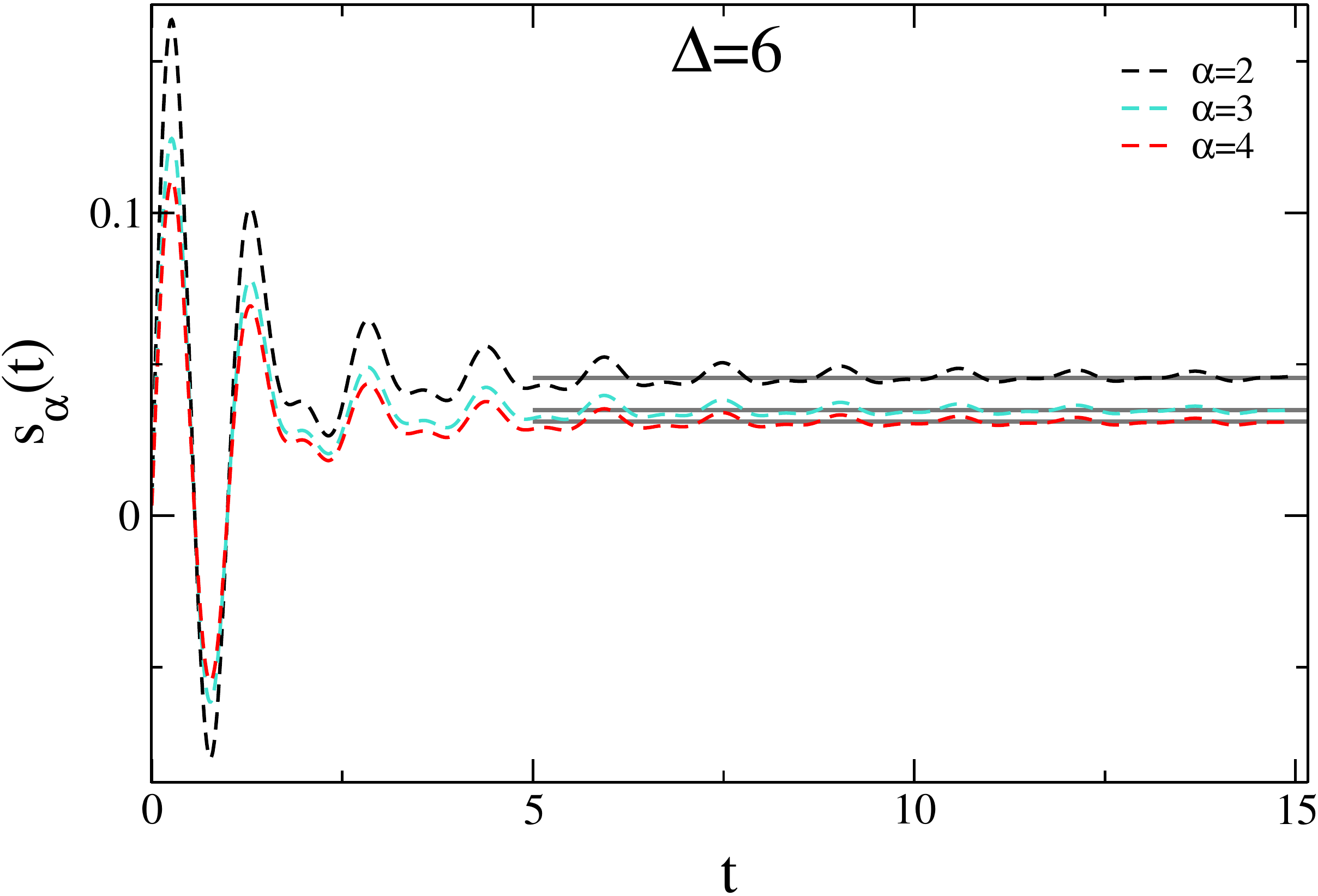}}
    \caption{\label{fig:Neel}
        Numerical results for instantaneous R\'enyi slope $s_{\alpha}(t)$
    after the quench from the N\'eel state~\eqref{eq:Neel}. The dashed lines represent
    the numerical data, with different colours distinguishing between different values
    of $\alpha$, while horizontal solid lines denote the analytical predictions for
    $s_{\alpha}$. Each panel corresponds to a different value of the anisotropy
    parameter: (a) $\Delta=2$, (b) $\Delta=3$, (c) $\Delta=4$, and (d) $\Delta=6$.
    }
\end{figure*}
Finally, let us consider the anisotropic spin-$1/2$ Heisenberg chain given by the Hamiltonian 
\be \label{eq:ham-xxz}
    H=\frac{J}{4}\sum_{j=1}^L\Big[\sigma^x_{j}\sigma^x_{j+1}+
    \sigma_{j}^y\sigma_{j+1}^y+\Delta(\sigma_{j}^z\sigma_{j+1}^z-1)\Big], 
\ee
where $\sigma_j^{x,y,z}$ are Pauli matrices acting at 
site $j$ and $\Delta$ is the anisotropy parameter, while the boundary conditions
are assumed to be periodic.
For the initial states of the quench protocol we consider the \emph{N\'eel state}
$\ket{\Psi_{\mathrm{N}}}$, and the \emph{Majumdar-Ghosh state} $\ket{\Psi_{\mathrm{MG}}}$,
defined as
\begin{align}
\label{eq:Neel}
    \ket{\Psi_{\mathrm{N}}}&=\frac{1}{\sqrt{2}}
    \Big(\ket{\uparrow\downarrow}^{\otimes L/2}+\ket{\downarrow\uparrow}^{\otimes L/2}\Big),\\
    \label{eq:MG}
    \ket{\Psi_{\mathrm{MG}}}&=\frac{1}{\sqrt{2}}\Big(\ket{\uparrow\downarrow}
    -\ket{\downarrow\uparrow}\Big)^{\otimes L/2}.
\end{align}
Since these states are integrable~\cite{piroli2017what}, we expect our result to
apply. Moreover, we can efficiently
characterise the late-time stationary state using the quench-action
approach~\cite{caux2016quench,pozsgay2018overlaps}.  Therefore, we are able to
explicitly evaluate the prediction given by Eqs.~\eqref{eq:convenientslope}
and~\eqref{eq:yTBA} (see Appendix~\ref{sec:tba-xxz} for additional details),
and compare it with numerical data obtained through 
the infinite time-evolving block decimation (iTEBD)~\cite{schollwock2011density}
method. See Appendix~\ref{sec:numerics} for the details on the implementation.

In particular, we evaluate the R\'enyi entanglement entropies between the two
halves of an infinite chain, and then express the \emph{instantaneous R\'enyi slope}
$s_{\alpha}(t)$, defined as the time-derivative of the R\'enyi entropy
$S^{(\alpha)}_{\rm half}(t)$
\be
s_{\alpha}(t)=\frac{{\rm d} S_{\rm half}^{(\alpha)}(t)}{{\rm d} t}.
\ee
Since the subsystem in question is half-infinite, we expect our
prediction to coincide with the instantaneous slope  in the $t\to\infty$ limit,
\be
\lim_{t\to\infty} s_{\alpha}(t) = s_{\alpha}.
\ee
This limit, however, cannot be accessed numerically because the linear growth
of entanglement after the quench implies exponential growth of computational
complexity to simulate the dynamics. Therefore, we have to compare the
prediction with finite-time data. We consider the regime $\Delta \ge 1$. In
particular, since it is well known that for the initial states~\eqref{eq:Neel}
and~\eqref{eq:MG} the entanglement slope increases when approaching $\Delta=1$
from above (see, e.g., Refs.~\cite{alba2017entanglement,alba2018entanglement}),
we restrict ourself to $\Delta>1$.

\begin{figure*}
    \centering
    (a)%
    \raisebox{-\totalheight+\baselineskip}[0pt][\totalheight]{\includegraphics[width=0.95\columnwidth]{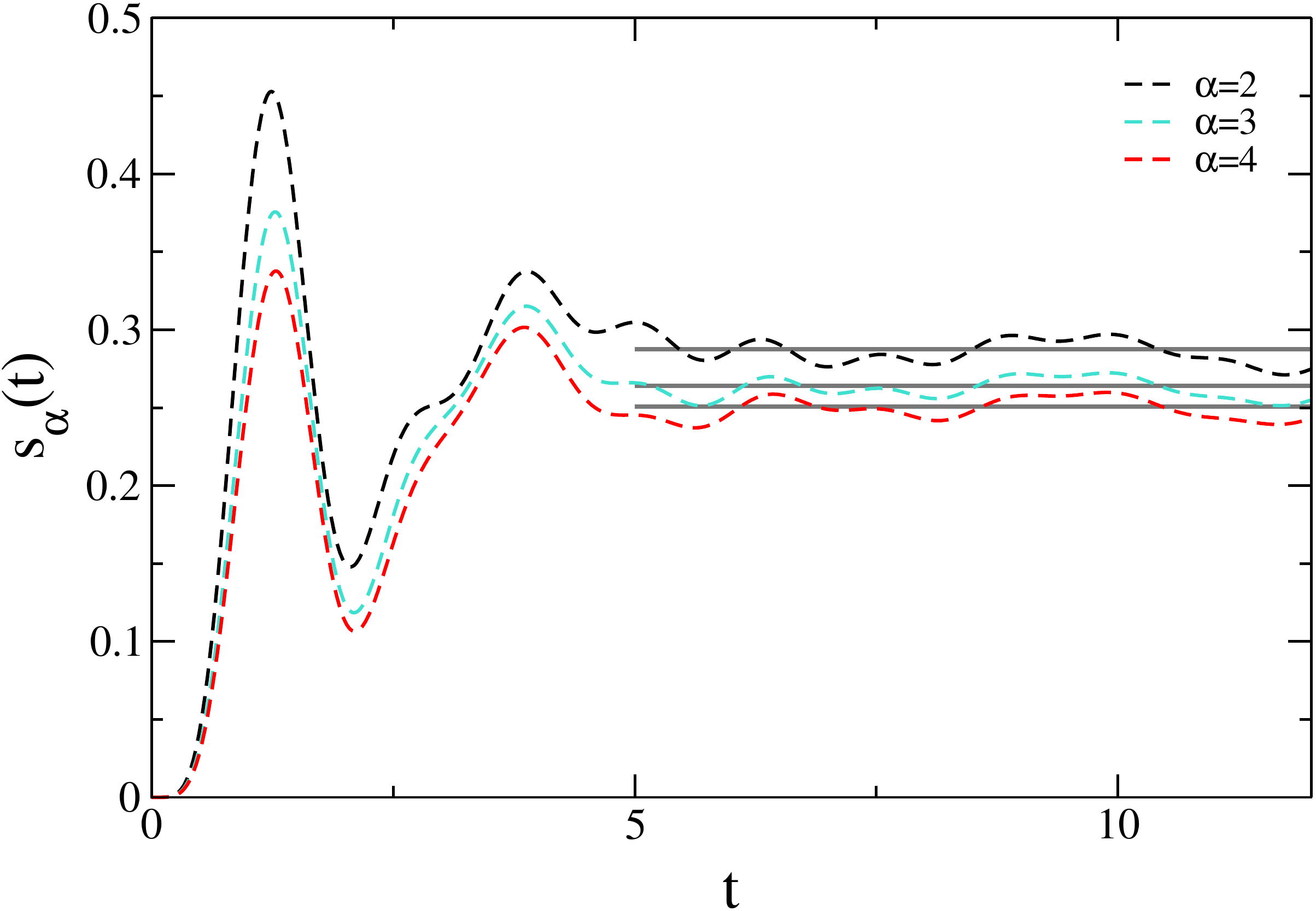}}\hfill
    (b)%
    \raisebox{-\totalheight+\baselineskip}[0pt][\totalheight]{\includegraphics[width=0.95\columnwidth]{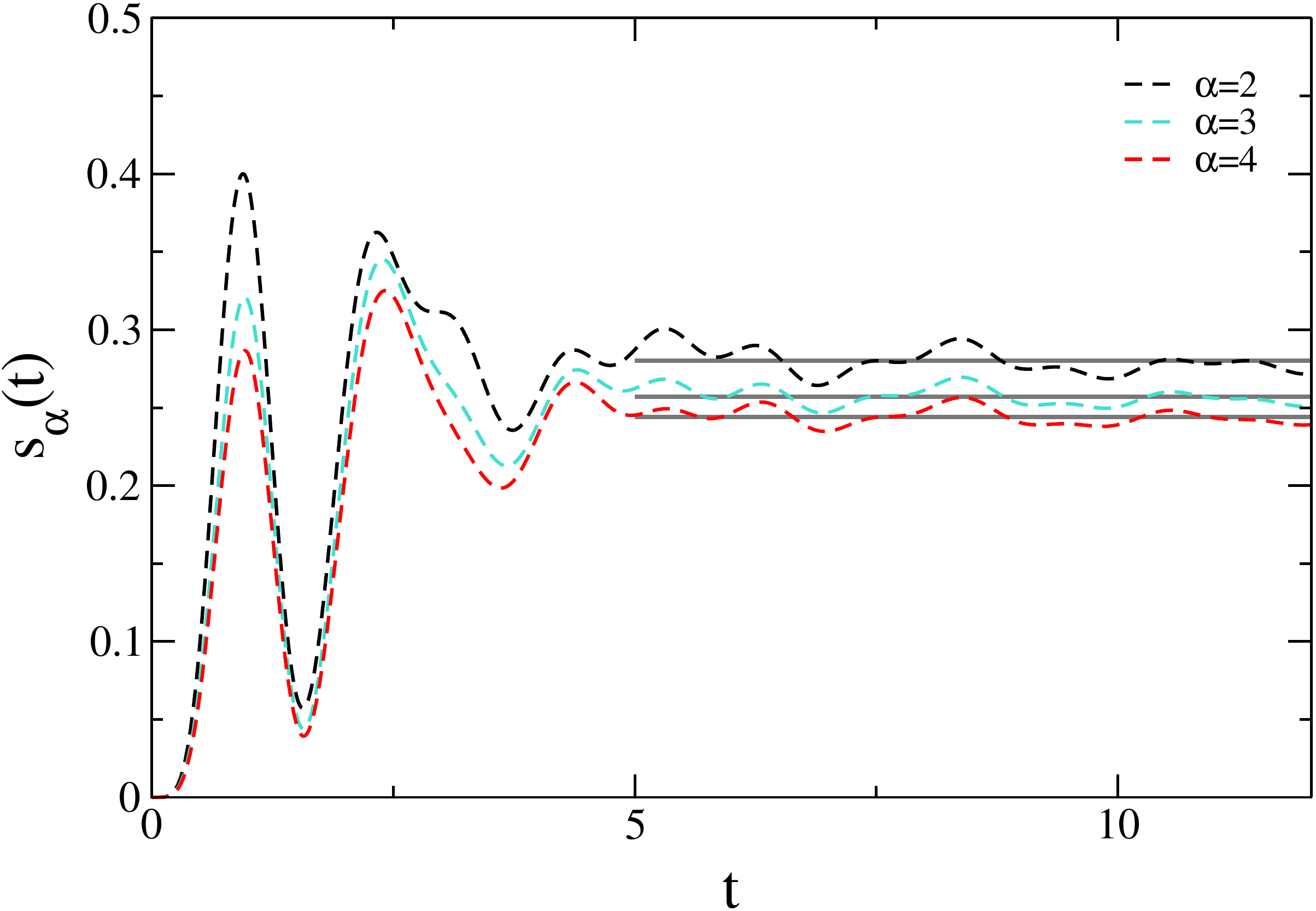}}
    \caption{\label{fig:MajumdarGhosh}
    Dynamics of the instantaneous R\'enyi slope $s_{\alpha}(t)$ after the quench from
    the Majumdar-Ghosh state~\eqref{eq:MG}. The dashed lines represent the numerical data,
    and solid lines denote the analytical predictions for the asymptotic slope $s_{\alpha}$.
    Different colours distinguish between different $\alpha$, while the value of the
    anisotropy parameter is $\Delta=3$ in the panel (a), and $\Delta=4$ in (b).
    }
\end{figure*}

The numerical results for the quench from the N\'eel state, and a range of different
values of $\Delta$, is shown in Fig.~\ref{fig:Neel}. We observe
that the data exhibit complicated non-universal dynamics at short times, and then
start approaching the asymptotic value. For all values of $\Delta$  and $\alpha$,
the agreement between the finite-time dynamics and the asymptotic prediction are
extremely good, considering the fact that the accessible times are relatively short.
Note that the seemingly larger deviations seen at $\Delta=2$ are due to long-wavelength damped oscillations (observed generically in integrable systems, see, e.g., Refs.~\cite{fagotti2008evolution,alba2017entanglement}). Similarly, in Fig.~\ref{fig:MajumdarGhosh} we test the prediction with the data for
the quench from the Majumdar-Ghosh state. The numerics again matches the asymptotic
slope very well.

\section{Implications for the Quasiparticle Picture}\label{sec:quasiparticle}
Here we generalise an argument presented in Ref.~\cite{klobas2021entanglement} for the case of Rule 54, to argue that the expressions \eqref{eq:convenientdensity} and \eqref{eq:convenientslope} \emph{cannot} be interpreted in terms of the quasiparticle picture of Ref.~\cite{calabrese2005evolution}. 
To explain our reasoning, let us begin by briefly recalling the essential ingredients of the latter. 

The quasiparticle picture is based on two basic postulates~\cite{calabrese2005evolution}: \begin{enumerate*}[label=(\roman*)] \item the initial state $\ket{\Psi_0}$ produces pairs of \emph{correlated} (or \emph{entangled}) quasiparticles --- objects propagating as free classical particles --- at every point in space; \item at any time $t\geq0$, the entanglement between a given subsystem $A$ and its complement $\bar A$ is proportional to the number of correlated pairs shared between the two.\end{enumerate*}

Admitting that, in general, quasiparticles can come in multiple species --- labelled by a positive integer $n$ --- and have a nontrivial dispersion relation --- parametrised by a rapidity $\mu$ --- the two postulates above lead to the following evolution equation for a given R\'enyi entropy
\be
    S_{A, \rm qp}^{(\alpha)}(t) =  \int_+ \!\!\! {{\rm d}\boldsymbol \mu}\, 
    \min(2 v_{\rm qp}(\boldsymbol \mu) t,|A|) s_{\alpha}(\boldsymbol \mu)\,.
\label{eq:quasiparticlemultiS}
\ee
Here we adopted the shorthand notation of Eqs.~\eqref{eq:shorthand} and \eqref{eq:yTBA}, and used that, for solvable initial states, the correlated pairs are formed by quasiparticles of the same species and opposite rapidity~\cite{alba2017entanglement}. Moreover, we denoted by 
\be
v_{\rm qp}(\boldsymbol \mu)=v_{n, \rm qp}(\mu),
\ee
the velocity of the quasiparticles of species $n$ and rapidity $\mu$, and by 
\be
s_{\alpha}(\boldsymbol \mu)=s_{n, \alpha}(\mu),
\ee
the contribution  to the R\'enyi entropy of a pair of quasiparticles of species $n$ and rapidities $\pm\mu$.

To make Eq.~\eqref{eq:quasiparticlemultiS} truly predictive one needs to specify $v_{\rm qp}(\boldsymbol \mu)$ and $s_{\alpha}(\boldsymbol \mu)$. In particular, Ref.~\cite{alba2017entanglement} showed that one can describe the dynamics of von-Neumann entropy by making the two following assumptions: \begin{enumerate*}[label=(\roman*)] \item $s_{1}(\boldsymbol \mu)$ is the density of entanglement entropy (cf.~\eqref{eq:YYentropydensity}); \item $v_{\rm qp}(\boldsymbol \mu)$ is given by velocity of excitations on the thermodynamic macrostate describing the stationary value of local observables after the quench. \end{enumerate*} The latter is fixed by Eqs.~\eqref{eq:totalrho} and \eqref{eq:vtotalrho}, where $\vartheta(\boldsymbol \mu)$ is the filling function of the relevant stationary state. 

Here we do not use these assumptions and, for the moment, we compare \eqref{eq:quasiparticlemultiS} to \eqref{eq:convenientdensity} and \eqref{eq:convenientslope} leaving $v_{\rm qp}(\boldsymbol \mu)$ and $s_{\alpha}(\boldsymbol \mu)$ unspecified. In particular, we consider an initial state producing a filling function of the form
\be
\vartheta_n(\mu)= \delta_{n,\bar n} \begin{cases}
    \vartheta & \mu \in [\bar{\mu}-\delta,\bar{\mu}+\delta],\\
    0 & \text{otherwise},
\end{cases}
\label{eq:testfilling}
\ee
with $\vartheta\leq1$ and $\delta\ll1$. In this case, we see that the three equations are compatible for all $\bar n$ and $\bar \mu$ only if
\be
v_{\rm qp}(\boldsymbol \mu) = \frac{\varepsilon'(\boldsymbol \mu)}{|p'(\boldsymbol \mu)|}\frac{\displaystyle\ln\left[(1- \vartheta(\boldsymbol \mu))^{\alpha} \!\!+  
    \frac{\vartheta(\boldsymbol \mu)^{\alpha}}{y_\alpha(\boldsymbol \mu)} \right]}{\displaystyle \ln\left[(1- \vartheta(\boldsymbol \mu))^{\alpha} \!\!+  
    \frac{\vartheta(\boldsymbol \mu)^{\alpha}}{x_\alpha(\boldsymbol \mu)} \right]}\!.
    \label{eq:conditionqp}
\ee
The crucial observation at this point is that the right hand side of \eqref{eq:conditionqp} depends nontrivially on $\alpha$. Therefore, one needs to allow for an $\alpha$-dependent quasiparticle velocity $v_{\rm qp}(\boldsymbol \mu)$. At first sight this might seem enough to exclude the applicability of any quasi-particle picture. Indeed it is natural to require that the properties of quasiparticles have to be fixed by initial state and dynamics and cannot depend on the specific observable (e.g.\ on $\alpha$). Here, however, we allow for more flexibility: since R\'enyi entropies have a non-linear dependence on the state of the system, their stationary values are described by an $\alpha$-dependent macrostate with filling function~\cite{alba2017quench}
\be
\vartheta_\alpha(\boldsymbol \mu) = \frac{\vartheta(\boldsymbol \mu)^\alpha}{x_\alpha(\boldsymbol \mu)(1-\vartheta(\boldsymbol \mu))^\alpha+\vartheta(\boldsymbol \mu)^\alpha},
\ee 
where $\vartheta(\boldsymbol \mu)$ is the filling function~\eqref{eq:testfilling}. One can then wonder whether the velocity of excitations on the $\alpha$-dependent macrostate --- obtained by solving Eqs.~\eqref{eq:totalrho} and \eqref{eq:vtotalrho} with $\vartheta(\boldsymbol \mu)$ --- coincides with~\eqref{eq:conditionqp}. However, this is the case only in the limit $\alpha\to1$.     

Since the velocity on the r.h.s.\ of~\eqref{eq:conditionqp} cannot be interpreted as the velocity of the excitations on a physically meaningful macrostate, we conclude that the quasiparticle picture does not describe the dynamics of R\'enyi entropies, at least at the quantitative level.

\begin{figure*}
    \centering
    (a)%
    \raisebox{-\totalheight+\baselineskip}[0pt][\totalheight]{\includegraphics[width=0.93\columnwidth]{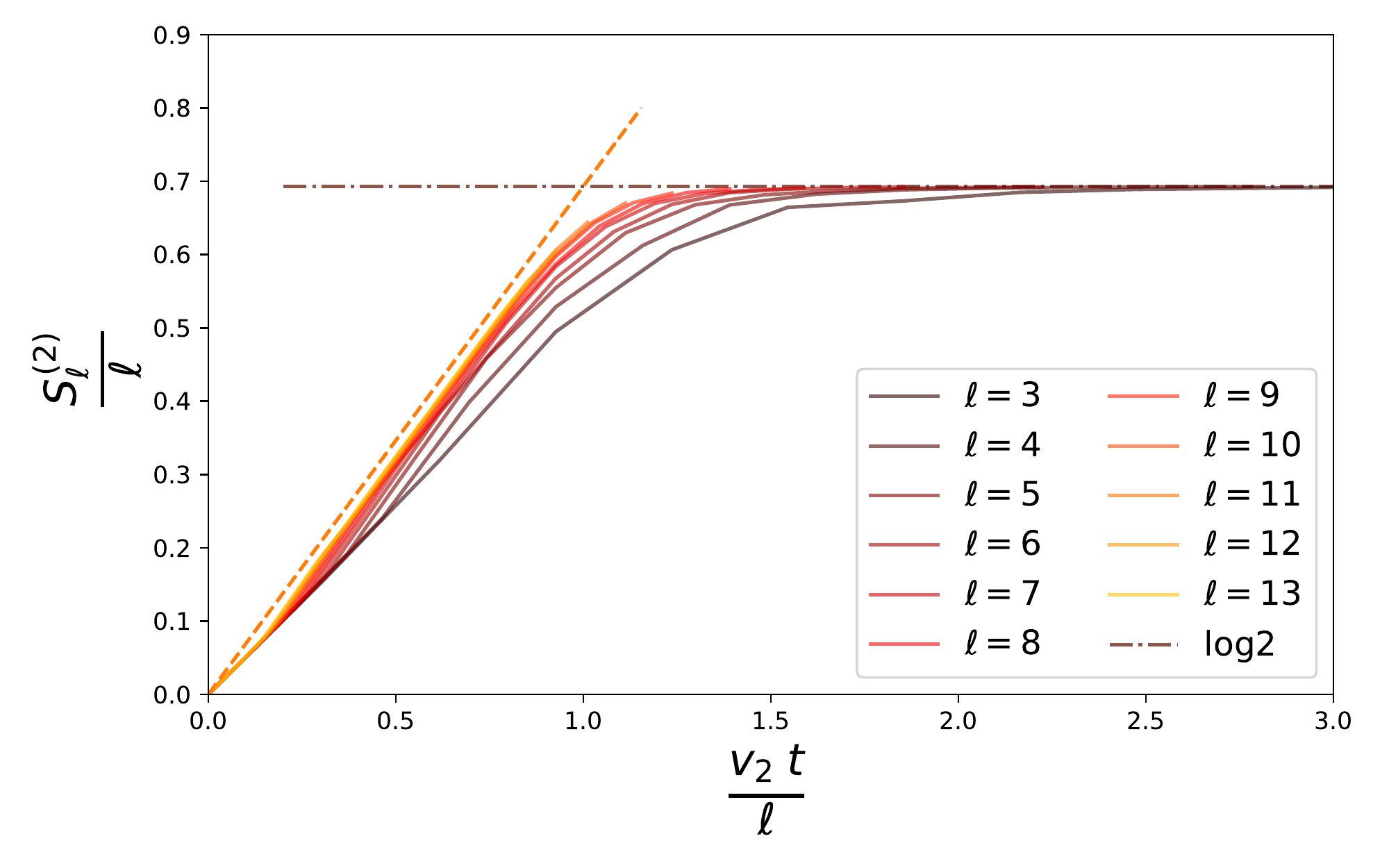}}\hfill
    (b)%
    \raisebox{-\totalheight+\baselineskip}[0pt][\totalheight]{\includegraphics[width=0.93\columnwidth]{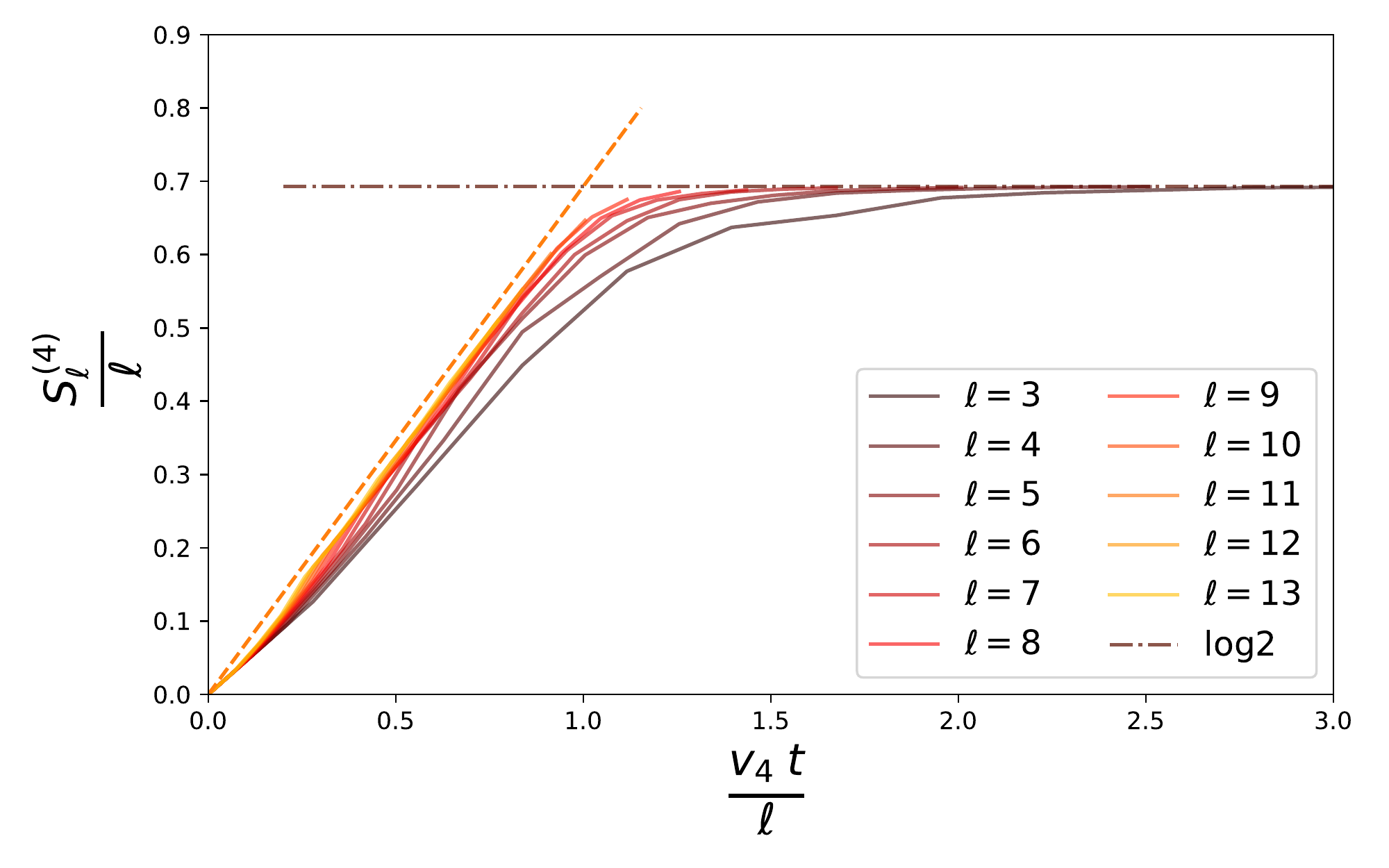}}
    \caption{Tensor network simulations of the entanglement dynamics for Rule 54. We consider an open system of $L=50$ sites and we focus on the subsystem $A$ made of the first $\ell$ sites, with $\ell=3,\dots 10$, represented by solid lines with different colours (see the legend). We fix the bond dimension equal to 4096. The dashed line is the exact prediction for  $s_\alpha t$ (note the absence of the factor of $2$ w.r.t.\ Eq.~\eqref{eq:quasiconj} due to the choice of different boundary conditions), while the dot-dashed line represents the asymptotic thermodynamic entropy $d_\alpha \ell$. Our conjecture \eqref{eq:quasiconj} corresponds to the joining of the two straight lines. The {\it entanglement velocity} that rescales the time is $v_{\alpha}={s_{\alpha}}/{d_{\alpha}}$. }
    \label{fig:my_label}
\end{figure*}

\section{Beyond the Linear Growth Regime}
\label{sec:finite}
One of the benefits of the quasiparticle picture is that, just using the few assumptions recalled in the previous section, one can quantitatively determine the evolution of entanglement in a wealth of different settings. Essentially, the dynamics of entanglement becomes a problem of one-dimensional kinematics: knowledge of velocities and entanglement contributions of each species of quasiparticles is enough to immediately determine the whole dynamics of the entanglement of a finite subsystem (cf.~\eqref{eq:quasiparticlemultiS}). The breakdown of the quasi-particle picture for $\alpha\neq 1$ completely changes the game. Determining the full curve $S^{(\alpha)}_A(t)$ becomes a highly non-trivial task for interacting integrable systems and our results for slope and density do not seem sufficient to achieve it. Here we present some evidence suggesting that, in fact, they might be enough. 

To this end we make two minimal assumptions: \begin{enumerate*}[label=(\roman*)] \item Each ``mode'' with quantum number $\boldsymbol \mu$ evolves independently; \item\label{item:abruptSat} For each mode the entanglement grows with fixed slope $2 s_{\alpha}(\boldsymbol \mu)$ until it abruptly saturates to $d_{\alpha}(\boldsymbol \mu) |A|$ (the factor of 2 comes from the fact that the subsystem has two edges).\end{enumerate*} A way to justify the assumption of abrupt saturation is to argue that modes behave as chaotic systems following the membrane picture~\cite{nahum2017quantum, zhou2020entanglement}. 

The two assumptions above lead to the following evolution equation for a given R\'enyi entropy
\be
    S_{A, \rm conj}^{(\alpha)}(t) =  \int_+ \!\!\! {{\rm d}\boldsymbol \mu}\, 
    \min(2s_{\alpha}(\boldsymbol \mu) t,d_{\alpha}(\boldsymbol \mu) |A|) \,,
\label{eq:quasiconj}
\ee
which we conjecture apply at the leading order for large $t$ and $|A|$. 

Testing \eqref{eq:quasiconj} numerically in a standard interacting integrable model, such as the XXZ spin chain, is very hard (see the simulations for ${\alpha=1}$ in Ref.~\cite{alba2017entanglement}): one cannot typically access its regime of validity in a sufficiently controlled manner to wash off all sub-leading corrections. However, it is instructive to test it for Rule 54. Indeed, since in that model there is a single mode for all $\alpha$'s, we can directly verify Assumption \ref{item:abruptSat}, without worrying about the subtle effects of the integration over $\boldsymbol \mu$. Our numerical results based on tensor network simulations 
are reported in Fig. \ref{fig:my_label}. We can clearly see that as $\ell$ increases the data approach our conjecture quite neatly, although we cannot exclude a different crossover close to the saturation point.

\section{Conclusions}
\label{sec:discussion}

In this paper we investigated the growth of entanglement after quantum quenches in quantum many-body systems by characterising the evolution of the R\'enyi entropies of a compact subsystem. In the cases of interest here these quantities exhibit a linear growth in time followed by saturation to their ``thermodynamic value'', i.e., their value in the steady state. We showed that one can generically express the initial slope of a given R\'enyi entropy as the density of entropy in a particular state of the dual system, i.e., the system obtained swapping the roles of space and time, cf.\ Eq.~\eqref{eq:slopetodensity}. The latter state is directly expressed in terms of the fixed points of the space transfer matrix~\cite{banuls2009matrix,muellerhermes2012tensor,hastings2015connecting} --- also known as influence matrices~\cite{lerose2021influence} ---, which characterise the evolution of local observables in the thermodynamic limit. 

In cases where the dual system can be interpreted as an isolated quantum many-body system --- for example for  \emph{dual-unitary quantum circuits}~\cite{bertini2019exact} --- the slope of a R\'enyi entropy is given by the \emph{density of entropy in the stationary state of the dual system}. Crucially, because of crossing symmetry, this is the case also for relativistic quantum field theories, provided that the direct swap of space and time is replaced by an appropriate analytic continuation. 

We used this observation to find a closed-form expression for the slope of R\'enyi entropies in integrable relativistic quantum field theories, going beyond what is currently achievable by known approaches such as form factor expansions~\cite{castroalvaredo2019entanglement,castroalvaredo2020entanglement,lencses2020relaxation,murciano2021post}. Moreover, we argued that this expression can be directly extended to all TBA integrable models. The most general form of our formula is  reported in Eq.~\eqref{eq:conjecture}. To support the validity of Eq.~\eqref{eq:conjecture} we showed that it reproduces the only known results for the slopes of R\'enyi entropies in an interacting integrable system~\cite{klobas2021entanglement}, the quasiparticle picture prediction in the von Neumann limit~\cite{alba2017entanglement}, and it reduces to the correct non-interacting limit~\cite{fagotti2008evolution, fagotti2008dinamica}. We also provided stringent numerical checks for several quenches in the XXZ spin-1/2 chain.
We then used Eq.~\eqref{eq:conjecture} to argue that the quasiparticle-picture does not describe the growth of R\'enyi entropies away from the von Neumann limit, and, finally, we proposed an extension of our formula away from the initial growth regime.

Our results have two significant merits. First, with Eq.~\eqref{eq:slopetodensity} we provided a direct relation between growth of entanglement in a given isolated quantum many-body system and the spatial scaling of stationary entanglement in its dual. This complements an analogous relation discovered in Ref.~\cite{ippoliti2021fractal}, for systems with edge decoherence. Second, with Eq.~\eqref{eq:conjecture} we solved the long standing open problem of computing the growth of R\'enyi entropies in interacting integrable models.

The results presented in this paper open many significant directions for future research. 
A direct question concerns the possibility of devising generalisations of our approach to treat other relevant quantities or describe more general settings. For instance, a recent point of interest in the research on quantum many-body dynamics is to understand how the entanglement is split among different symmetry sectors, with symmetry resolved entropies directly measured in experiments~\cite{vitale2022symmetry}. While the quasiparticle picture has been shown to hold for free systems~\cite{parez2021quasiparticle, parez2021exact}, no result is available in the interacting case. An immediate question is then whether one can use the recently obtained explicit results for equilibrium states~\cite{piroli2022thermodynamic} to generalise our approach and access the full dynamics of symmetry resolved R\'enyi entropies. At the same time, it is also interesting to wonder whether our approach can be extended to inhomogeneous settings. Indeed, in this case the late-time quasi-stationary regime is still characterised using integrability via the framework of generalised hydrodynamics (GHD)~\cite{castroalvaredo2016emergent,bertini2016transport}, and an appropriate modification of the quasiparticle picture correctly characterises the time evolution of the von-Neumann entanglement entropy~\cite{bertini2018entanglement, alba2019entanglement}. 

A second set of questions, instead, stems from our findings on the inapplicability of the quasiparticle picture to describe R\'enyi entropies in interacting integrable models. Indeed, as touched upon in the introduction, the fact that the entanglement is propagated by quasiparticles --- rather than behaving as a membrane in the spacetime --- has direct consequences on its phenomenology. These are revealed, for instance, in the qualitative behaviour of the bipartite entanglement between a disjoint region and the rest of the system~\cite{asplund2015entanglement, nahum2017quantum, bertini2018entanglement, alba2019quantum}, or in a system of finite size~\cite{nahum2017quantum, bertini2018entanglement, modak2020entanglement}. Recent studies, however, suggest that this might be the case also concerning multipartite entanglement, which is conveniently characterised by the \emph{entanglement negativity} and the higher moments of the partial transpose of the reduced density matrix~\cite{peres1996separability,eisert1999comparison,horodecki2001separability,vidal2002computable,plenio2005logarithmic}
. Using the quasiparticle picture, Ref.~\cite{alba2019quantuminformation} argued that, after a quantum quench, the logarithmic negativity coincides with half of the R\'enyi mutual information with $\alpha=1/2$ --- a similar statement holds for the higher moments~\cite{murciano2021quench}. On the other hand, Ref.~\cite{bertini2022dynamics} has recently shown that in general this relation holds only in the early time regime. Therefore, it is interesting to wonder what happens for interacting integrable systems beyond this regime. Another important question relates to the possibility of developing an alternative emergent picture to describe the growth of entanglement in interacting integrable models. Since these systems are ultimately defined by the presence of stable quasiparticles at all energy scales, it is reasonable to expect that a more complicated quasiparticle picture --- perhaps involving quasiparticles propagating in the multi-replica space --- will be able to account for the growth of R\'enyi entropies.


Finally, we mention that our formula \eqref{eq:conjecture} for the slope of R\'enyi entropies in all TBA-integrable models has not been rigorously proven here, even though the arguments we provided leave little doubts on its validity. Nevertheless, the correspondence that we established between slope in the original model and steady state entropy in the dual model provides an ideal starting point for such a rigorous proof. A direct question for future research is then to devise such a rigorous proof --- for instance using the framework of algebraic Bethe ansatz.  Besides the major interest that such a proof would have \emph{per se}, it would also lead to a rigorous validation of the quasiparticle conjecture in the replica limit $\alpha\to1$ --- a problem that has been open since 2005~\cite{calabrese2005evolution}.

\acknowledgments
This work has been supported by the Royal Society through the University Research Fellowship No.\ 201101 (BB), by the EPSRC under grant EP/S020527/1 (KK), by ERC under Consolidator grant NEMO 771536 (GL, PC). VA, BB and KK thank SISSA for hospitality in the early stage of this project.

\appendix

\section{Fixed points as stationary density matrices}\label{sec:fixedPointsCircuits}
The fact that the matrices $M_{\mathrm{L/R},t}$ exhibit the Cholesky
decomposition~\eqref{eq:positivityMLMR} follows directly from the explicit form of 
the fixed points. In particular, the fixed points of the circuit shown in Fig.~\ref{fig:QC}
take the following form (see, e.g.,~\cite{muller2012tensor,klobas2021exactrelaxation})
\be
\begin{tikzpicture} [baseline={([yshift=-2.125ex]current bounding box.center)},scale=0.4]
    \draw[thick,gray,rounded corners=1.5,fill=gray,fill opacity=0.2] (-0.75,0) -- (5,0)
    -- (5,1) -- (0.25,5.75) -- (-0.75,5.75) -- cycle;
    \node at (0,6.125) {\scalebox{0.9}{$m_{\mathrm{R},t}$}};
    \node at (-2.375,0) {\scalebox{0.775}{$\ket{M_{\mathrm{R},t}}=$}};
    \draw[semithick,colLines,rounded corners=3.75] (3.25,0.25) -- (4.5,1.5) -- (5.125,1.5) 
    -- (5.125,-1.5) -- (4.5,-1.5) -- (3.25,-0.25);
    \draw[semithick,colLines,rounded corners=4.125] (1.25,0.25) -- (3.5,2.5) -- (5.25,2.5) 
    -- (5.25,-2.5) -- (3.5,-2.5) -- (1.25,-0.25);
    \draw[semithick,colLines,rounded corners=4.125] (-1,0.5) -- (-0.5,0.5) -- (2.5,3.5) -- 
    (5.375,3.5) -- (5.375,-3.5) -- (2.5,-3.5) -- (-0.5,-0.5) -- (-1,-0.5);
    \draw[semithick,colLines,rounded corners=4.125] (-1,2.5) -- (-0.5,2.5) -- (1.5,4.5) -- 
    (5.5,4.5) -- (5.5,-4.5) -- (1.5,-4.5) -- (-0.5,-2.5) -- (-1,-2.5);
     \draw[semithick,colLines,rounded corners=4.125] (-1,4.5) -- (-0.5,4.5) -- (0.5,5.5) -- 
    (5.625,5.5) -- (5.625,-5.5) -- (0.5,-5.5) -- (-0.5,-4.5) -- (-1,-4.5);

    \draw[semithick,colLines,rounded corners=4.125] (-1,5.5) -- (-0.5,5.5) -- (4.75,0.25);
    \draw[semithick,colLines,rounded corners=4.125] (-1,3.5) -- (-0.5,3.5) -- (2.75,0.25);
    \draw[semithick,colLines,rounded corners=4.125] (-1,1.5) -- (-0.5,1.5) -- (0.75,0.25);
    \draw[semithick,colLines,rounded corners=4.125] (-1,-5.5) -- (-0.5,-5.5) -- (4.75,-0.25);
    \draw[semithick,colLines,rounded corners=4.125] (-1,-3.5) -- (-0.5,-3.5) -- (2.75,-0.25);
    \draw[semithick,colLines,rounded corners=4.125] (-1,-1.5) -- (-0.5,-1.5) -- (0.75,-0.25);

    \prop{0}{1}{colU}
    \prop{0}{3}{colU}
    \prop{0}{5}{colU}
    \prop{1}{2}{colU}
    \prop{1}{4}{colU}
    \prop{2}{1}{colU}
    \prop{2}{3}{colU}
    \prop{3}{2}{colU}
    \prop{4}{1}{colU}
    \prop{0}{-1}{colUc}
    \prop{0}{-3}{colUc}
    \prop{0}{-5}{colUc}
    \prop{1}{-2}{colUc}
    \prop{1}{-4}{colUc}
    \prop{2}{-1}{colUc}
    \prop{2}{-3}{colUc}
    \prop{3}{-2}{colUc}
    \prop{4}{-1}{colUc}
    \istate{0.75}{0.25}{colIst}
    \istate{1.25}{0.25}{colIst}
    \istate{2.75}{0.25}{colIst}
    \istate{3.25}{0.25}{colIst}
    \istate{4.75}{0.25}{colIst}
    \istate{0.75}{-0.25}{colIstC}
    \istate{1.25}{-0.25}{colIstC}
    \istate{2.75}{-0.25}{colIstC}
    \istate{3.25}{-0.25}{colIstC}
    \istate{4.75}{-0.25}{colIstC}
\end{tikzpicture}
\,,
\begin{tikzpicture} [baseline={([yshift=-2.6ex]current bounding box.center)},scale=0.4]
    \draw[thick,gray,rounded corners=1.5,fill=gray,fill opacity=0.2] (3.75,0) -- (-1,0)
    -- (-1,1) -- (3.25,5.75) -- (3.75,5.75) -- cycle;
    \node at (3.25,6.125) {\scalebox{0.9}{$m_{\mathrm{L},t}$}};
    \node at (-3,0) {\scalebox{0.775}{$\bra{M_{\mathrm{L},t}}=$}};
    \draw[semithick,colLines,rounded corners=3.75] (0.75,0.25) -- (-0.5,1.5) -- (-1.125,1.5) -- 
    (-1.125,-1.5) -- (-0.5,-1.5) -- (0.75,-0.25);
    \draw[semithick,colLines,rounded corners=4.125] (2.75,0.25) -- (0.5,2.5) -- (-1.25,2.5) -- 
    (-1.25,-2.5) -- (0.5,-2.5) -- (2.75,-0.25);
    \draw[semithick,colLines,rounded corners=4.125] (4,1.5) -- (3.5,1.5) -- (3,2) -- (1.5,3.5) 
    -- (-1.375,3.5) -- (-1.375,-3.5) -- (1.5,-3.5) -- (3,-2) -- (3.5,-1.5) -- (4,-1.5);
    \draw[semithick,colLines,rounded corners=4.125] (4,3.5) -- (3.5,3.5) -- (3,4) -- (2.5,4.5) 
    -- (-1.5,4.5) -- (-1.5,-4.5) -- (2.5,-4.5) -- (3,-4) -- (3.5,-3.5) -- (4,-3.5);
    \draw[semithick,colLines,rounded corners=4.125] (4,5.5) -- (-1.625,5.5) -- (-1.625,-5.5) 
    -- (4,-5.5);

    \draw[semithick,colLines,rounded corners=4.125] (-0.75,0.25) -- (3.5,4.5) -- (4,4.5);
    \draw[semithick,colLines,rounded corners=4.125] (1.25,0.25) -- (3.5,2.5) -- (4,2.5);
    \draw[semithick,colLines,rounded corners=4.125] (3.25,0.25) -- (3.5,0.5) -- (4,0.5);
    \draw[semithick,colLines,rounded corners=4.125] (-0.75,-0.25) -- (3.5,-4.5) -- (4,-4.5);
    \draw[semithick,colLines,rounded corners=4.125] (1.25,-0.25) -- (3.5,-2.5) -- (4,-2.5);
    \draw[semithick,colLines,rounded corners=4.125] (3.25,-0.25) -- (3.5,-0.5) -- (4,-0.5);

    \prop{0}{1}{colU}
    \prop{1}{2}{colU}
    \prop{2}{1}{colU}
    \prop{2}{3}{colU}
    \prop{3}{2}{colU}
    \prop{3}{4}{colU}
    \prop{0}{-1}{colUc}
    \prop{1}{-2}{colUc}
    \prop{2}{-1}{colUc}
    \prop{2}{-3}{colUc}
    \prop{3}{-2}{colUc}
    \prop{3}{-4}{colUc}
    \istate{-0.75}{0.25}{colIst}
    \istate{0.75}{0.25}{colIst}
    \istate{1.25}{0.25}{colIst}
    \istate{2.75}{0.25}{colIst}
    \istate{3.25}{0.25}{colIst}
    \istate{-0.75}{-0.25}{colIstC}
    \istate{0.75}{-0.25}{colIstC}
    \istate{1.25}{-0.25}{colIstC}
    \istate{2.75}{-0.25}{colIstC}
    \istate{3.25}{-0.25}{colIstC}
\end{tikzpicture}
\mkern-6mu.\mkern-4mu
\ee
Here we introduced the operators $m_{\mathrm{R},t}$ and $m_{\mathrm{L},t}$ (in grey),
which map from right to left. Recalling now that the corresponding matrices
$M_{\mathrm{R}/\mathrm{L},t}$ act on the $2t$ horizontal legs at the bottom,
and map them to the $2t$ legs at the top, we can immediately express
them in terms of $m_{\mathrm{R},t}$, $m_{\mathrm{L},t}$ as
\be
M_{\mathrm{R},t} = m_{\mathrm{R},t}^{\phantom{\dagger}}m_{\mathrm{R},t}^{\dagger},\qquad
M_{\mathrm{L},t} = m_{\mathrm{L},t}^{\dagger}m_{\mathrm{L},t}^{\phantom{\dagger}}.
\ee
Note that we only considered the case with the initial state in the product
form. With minor modifications, however, the argument can be repeated also for
the initial state in the form of a matrix-product state (MPS), as long as the
MPS transfer matrix has a unique dominant eigenvector.

\section{Partially decoupled form of \eqref{eq:yTBA} and \eqref{eq:xTBA}}
In systems with multiple types of particle species, Eqs.~\eqref{eq:yTBA}
and~\eqref{eq:xTBA} involve both an integral over rapidities, and an infinite
sum over the particle species. However, using standard TBA
manipulations~\cite{takahashi1999thermodynamics} the equations can be put in an
equivalent form, referred to as the \emph{decoupled form}, so that each
particle species $n$ is only coupled to $n+1$, and $n-1$, which makes the set
of equations simpler to solve.

For simplicity we restrict the discussion to the systems with the \emph{even} kernel 
that is in the difference form,
\be
T_{nm}(\lambda,\mu)=T_{nm}(\lambda-\mu)=T_{nm}(\mu-\lambda),
\ee
but a similar manipulation could be done more generally. In this case, the
integral equations~\eqref{eq:xTBA} and~\eqref{eq:yTBA} can be succinctly
expressed as
\begin{align}
\ln x_{n}(\lambda)
    \mkern-4mu&=\mkern-8mu\sum_{m}\mkern-4mu\Big(\mkern-3mu
T_{nm}\ast
\ln\mkern-6mu\big[(1-\vartheta_{m})^\alpha
\mkern-5mu+\mkern-4mu\frac{\vartheta_m^{\alpha}}{x_{m}}
\big]\!\Big)(\lambda),
    \label{eq:equivalentXYtba}\\
\notag
\ln y_{n}(\lambda)
    \mkern-4mu&=\mkern-8mu\sum_{m}\mkern-4mu\Big(\mkern-3mu
T_{nm}\ast
\mathrm{sgn}(\cdot)\ln\mkern-6mu\big[(1-\vartheta_{m})^\alpha
\mkern-5mu+\mkern-4mu\frac{\vartheta_m^{\alpha}}{y_{m}^{\mathrm{sgn}(\cdot)}}
\big]\!\Big)(\lambda),
\end{align}
where we introduced the shorthand notation $\ast$ for the convolution,
\be
(f\ast g)(\lambda) = \int \!\!{{\rm d}\mu} f(\lambda-\mu) g(\mu),
\ee
and $\mathrm{sgn}(\cdot)$ is the sign function,
\be
\mathrm{sgn}(x)=\begin{cases}
    1,\qquad &x>0,\\
    0,\qquad &x=0,\\
    -1,\qquad &x<0.
\end{cases}
\ee
Note that we dropped the explicit dependence on $\alpha$ from $x_{\alpha,n}$ and
$y_{\alpha,n}$ to ease the notation.

A necessary ingredient for the decoupled form is the existence of the function
$s(\lambda)$ that satisfies the following set of equations,
\be
\begin{aligned}
    T_{1m}(\lambda) &=  s \ast T_{2,m}(\lambda)  + \delta_{2,m} s(\lambda)\\
    T_{nm}(\lambda) &=  s \ast (T_{n-1,m}+T_{n+1,m})(\lambda)  \\
    &+ (\delta_{n-1,m}+ \delta_{n+1,m}) s(\lambda), \qquad n\geq 2\,.
\end{aligned}
\ee
Combining this with~\eqref{eq:equivalentXYtba}, we finally obtain the following
equivalent form of~\eqref{eq:xTBA}, and~\eqref{eq:yTBA},
\be
\begin{aligned}
    \ln x_{1}(\lambda) &=  s \ast 
    \ln\left[{(1-\vartheta_{2})^\alpha x_{2} +\vartheta_{2}^\alpha}\right](\lambda),\\
    \ln x_{n}(\lambda)  &= s \ast \ln
    \left[{(1-\vartheta_{n-1})^\alpha x_{n-1} +\vartheta_{n-1}^\alpha}\right]\!\!(\lambda)\\
    &+ s \ast 
    \ln\left[{(1-\vartheta_{n+1})^\alpha x_{n+1} 
    +\vartheta_{n+1}^\alpha}\right]\!(\lambda),\mkern-50mu\\
    \mkern-50mu\ln y_{1}(\lambda) &=  s \ast \mathrm{sgn}(\cdot) 
    \ln \left[{(1-\vartheta_{2})^\alpha y_{2}^{\mathrm{sgn}(\cdot)}\!\!
    +\vartheta_{2}^\alpha}\right](\lambda),\mkern-50mu\\
    \mkern-50mu\ln y_{n}(\lambda)  &= s \ast \mathrm{sgn}(\cdot) 
    \ln\left[{(1-\vartheta_{n-1})^\alpha y_{n-1}^{\mathrm{sgn}(\cdot)} 
    +\vartheta_{n-1}^\alpha}\right](\lambda)\mkern-50mu\\
    &+s \ast \mathrm{sgn}(\cdot) 
    \ln\left[{(1-\vartheta_{n+1})^\alpha  y_{n+1}^{\mathrm{sgn}(\cdot)}
    +\vartheta_{n+1}^\alpha}\right](\lambda).\mkern-50mu
\end{aligned}
\ee

\section{TBA equations for XXZ}\label{sec:tba-xxz}
Here we summarise the relevant details of the TBA description of the XXZ model
in the $\Delta>1$ regime~\cite{takahashi1999thermodynamics}. 
It is convenient to parametrise $\Delta$ as 
\be
\Delta=\cosh(\eta),\qquad \eta>0,
\ee
and express physically relevant quantities (such as energy and momentum) in terms of $\eta$
rather than $\Delta$.

The elementary excitations (magnons) can form infinitely many different
bound states, while their rapidity is constrained to $\abs{\lambda}\le {\pi}/{2}$,
therefore the integral over ${\boldsymbol \lambda}=(\lambda,n)$ should be understood
as
\be
\int d {\boldsymbol \lambda}=\sum_{n=1}^{\infty}\int_{-\frac{\pi}{2}}^{\frac{\pi}{2}} d\lambda,
\ee
and in particular integrals over \emph{positive} rapidities are meant as
\be
\int_{+} d {\boldsymbol \lambda}=\sum_{n=1}^{\infty}\int_{0}^{\frac{\pi}{2}} d\lambda.
\ee
Derivatives of bare energy $\varepsilon_n^{\prime}(\lambda)$ and momentum 
$p_n^{\prime}(\lambda)$ read as
\be
\begin{aligned}
    \frac{1}{2 \pi}\varepsilon_n^{\prime}(\lambda)&=
\frac{1}{\pi}
    \frac{\sin(2 \lambda)\sinh(\eta)\sinh(n \eta)}{\big(\cosh(n \eta)-\cos(2 \lambda)\big)^2},\\
    \frac{1}{2\pi}{p_n^{\prime}(\lambda)}&=
    \frac{1}{\pi}\frac{\sinh(n\eta)}{\cosh(n \eta)-\cos(2 \lambda)},
\end{aligned}
\ee
while the kernel takes a difference form,
$T({\boldsymbol \lambda},{\boldsymbol \mu})=T_{nm}(\lambda-\mu)$, and is given by
\be
\begin{aligned}
    T_{nn}(\lambda)&=
\frac{1}{2\pi}\smashoperator{\sum_{k=1}^{n}}
{{p^{\prime}_{2k}(\lambda)}},\\
    T_{nm}(\lambda)&=
\frac{1}{2\pi}\mkern16mu\smashoperator{\sum_{k=0}^{\frac{n+m-\abs{n-m}}{2}}}
{{p^{\prime}_{\abs{n-m}+2k}(\lambda)}},\quad m\neq n.
\end{aligned}
\ee
The last ingredient needed for the evaluation of the predictions is the filling 
function $\vartheta_{n}(\lambda)$. This is obtained
as~\cite{brockmann2014quench,pozsgay2018overlaps}
\be
\vartheta_{n}(\lambda)=\frac{1}{1+\eta_{n}(\lambda)},
\ee
where $\eta_{n}(\lambda)$ is the solution to the following integral equation
\be
\begin{aligned}
    \ln\eta_{n}(\lambda)&=g_{n}(\lambda)\\
    &+ \sum_{m=1}^{\infty} \smashoperator{\int_{-\,\frac{\pi}{2}}^{\frac{\pi}{2}}}
{\rm d}\mu\, T_{nm}(\mu-\lambda) \ln(1+\frac{1}{\eta_{m}(\mu)}).
\end{aligned}
\ee
Here, $g_{n}(\lambda)$ encodes the information about the initial state. 
For all integrable initial states the values of $g_{n}(\lambda)$ for
$n>1$ are expressed in terms of $g_1(\lambda)$ as 
\be
g_{n}(\lambda)=\sum_{k=1}^n g_1\!\!\left(\lambda+i\eta\frac{n+1-2k}{2}\right),
\ee
while $g_1(\lambda)$ for the two cases considered here reads as
\be
\begin{aligned}
    g_{1}^{\text{(MG)}}(\lambda)&=-\ln\left(
\frac{\sinh^4(\lambda)\cot^2(\lambda)}{\sin(2\lambda+i\eta)\sin(2\lambda-i\eta)}\right),\\
    g_{1}^{\text{(N\'eel)}}(\lambda)&=
    \frac{\tan\left(\lambda+\frac{i \eta}{2}\right)\tan\left(\lambda-\frac{i\eta}{2}\right)}
    {4\sin^2(2\lambda)}.
\end{aligned}
\ee
\section{Details on the iTEBD simulations}
\label{sec:numerics}
To perform the simulations we first build the MPS
representation for the initial state. Both the  N\'eel state and the
Majumdar-Ghosh state (cf.~\eqref{eq:Neel} and~\eqref{eq:MG}) admit a MPS
representation with small bond dimension. Then we perform the dynamics by
applying a second order Trotter decomposition of the time-evolution operator.
We verified that a Trotter step $\delta t=0.1$ is sufficient to ensure
time-converged results.  Due to the linear growth of entanglement the bond
dimension of the MPS representing the time-evolved state increases
exponentially with time. For this reason, at each step of the evolution we
perform a truncation of the MPS using singular value decomposition keeping the
largest  $\chi_\mathrm{max}$ singular values. To monitor the loss of precision,
we perform iTEBD simulations with increasing bond dimension $\chi_\mathrm{max}$
up to $\chi_\mathrm{max}=8192$ for $\Delta=2$, and $\chi_{\mathrm{max}}=4096$
for other values of $\Delta$. We then compare the data with two consecutive
values of $\chi_\mathrm{max}$ and only keep the data for which the two
simulations agree. This allows us to reach times of the order $t\lesssim 15$.

\bibliography{./bibliography}
\end{document}